\documentclass[journal]{IEEEtran}

\usepackage{amsmath,amsfonts}
\usepackage{algorithmic}
\usepackage{algorithm}
\usepackage{array}
\usepackage[caption=false,font=normalsize,labelfont=sf,textfont=sf]{subfig}
\usepackage{textcomp}
\usepackage{stfloats}
\usepackage{url}
\usepackage{verbatim}
\usepackage[table]{xcolor}
\usepackage{graphicx}
\usepackage{cite}
\usepackage{xcolor}
\usepackage{multirow}
\usepackage{adjustbox}
\usepackage{rotating}
\usepackage{pdflscape}
\usepackage{booktabs}
\usepackage{hyperref} 

\begin{document}
\bstctlcite{IEEEexample:BSTcontrol}
    \title{Graph Neural Networks and Reinforcement Learning for Proactive Application Image Placement}
 \author{Antonios~Makris,
        Theodoros~Theodoropoulos,
        Evangelos~Psomakelis,
        Emanuele~Carlini,
        Matteo~Mordacchini,
        Patrizio~Dazzi,
        and~Konstantinos~Tserpes

\IEEEcompsocitemizethanks{\IEEEcompsocthanksitem A. Makris, T. Theodoropoulos, E. Psomakelis and K.Tserpes are with the Department of Informatics and Telematics, Harokopio University of Athens, Greece and also with the School of Electrical and Computer Engineering, National Technical University of Athens. E-mail: \{amakris, ttheod, vpsomak, tserpes\}@hua.gr. 
\IEEEcompsocthanksitem E. Carlini and M. Mordacchini are with the Institute of Information Science and Technologies, National Research Council (CNR), Pisa, Italy. E-mail: \{emanuele.carlini, matteo.mordacchini\}@isti.cnr.it 
\IEEEcompsocthanksitem P. Dazzi is with the Department of Computer Science, University of Pisa, Italy. E-mail: patrizio.dazzi@unipi.it. 
}
\thanks{}
}


\maketitle

\begin{abstract}
The shift from Cloud Computing to a Cloud-Edge continuum presents new opportunities and challenges for data-intensive and interactive applications. 
Edge computing has garnered a lot of attention from both industry and academia in recent years, emerging as a key enabler for meeting the increasingly strict demands of Next Generation applications. In Edge computing the computations are placed closer to the end-users, to facilitate low-latency and high-bandwidth applications and services.
However, the distributed, dynamic, and heterogeneous nature of Edge computing, presents a significant challenge for service placement.
A critical aspect of Edge computing involves managing the placement of applications within the network system to minimize each application’s runtime, considering the resources available on system devices and the capabilities of the system's network.
The placement of application images must be proactively planned to minimize image transfer time, and meet the strict demands of the applications. 
In this regard, this paper proposes an approach for proactive image placement that combines Graph Neural Networks and actor-critic Reinforcement Learning, which is evaluated empirically and compared against various solutions. The findings indicate that although the proposed approach may result in longer execution times in certain scenarios, it consistently achieves superior outcomes in terms of application placement.
\end{abstract}


%
\IEEEpeerreviewmaketitle

\section{Introduction}
\label{introduction}
\IEEEPARstart{T}{he} rapid growth of data-intensive applications and latency-sensitive services (also referred as NextGen applications) is driving the adoption of edge computing architectures~\cite{makris2021cloud,korontanis2020inter,theodoropoulos2022cloud, 9524099}. Next generation applications involving data analytics, artificial intelligence, augmented/virtual reality, and Internet of Things (IoT) platforms require real-time processing and low latency to provide responsive and immersive user experiences. However, centralized cloud computing models often fail to meet these demands due to constraints around bandwidth, network latency, location-awareness, mobility support, security, and privacy.
Edge computing aims to address these challenges by enabling computation and data storage at the network edge, close to end users and data sources. Rather than sending all data to distant cloud servers, edge computing facilitates localized analysis and decision making using (resource-constrained) edge devices \cite{makris2022towards}. This allows latency-critical tasks to be performed within milliseconds and alleviates issues of high bandwidth consumption, intermittent connectivity, and single points of failure associated with cloud-only architectures.

A prevalent design pattern for NextGen edge applications involves microservice-based architectures that leverage lightweight and modular, independently deployable services~\cite{alam2018orchestration,makris2022transition}. By decomposing applications into granular, loosely-coupled microservices, developers gain flexibility to rapidly update individual components without disrupting others. This also enables services to be dynamically instantiated on demand based on real-time user presence, workload shifts and resource availability.
However, edge infrastructures introduce unique challenges for deploying and managing microservices at scale. Edge nodes have constrained storage, processing, and network resources compared to data centers. Additionally, the dynamic nature of edge workloads demands fast and adaptive service provisioning. Naively downloading container/VM images for each service instance activation from centralized repositories fails to meet these requirements.

In this work, our aim is to address this issue through a distributed edge image management approach. We propose partitioning edge nodes into cooperative groups and replicating microservice images across group members to minimize fetching latency. We formulate this optimization problem as a minimum vertex cover and present GNOSIS, a learning approach based
on the combination of Graph Neural Networks (GNNs) and Deep
Reinforcement Learning (DRL). This solution combines the representation power of a Graph Neural Network approach
with the ability of actor-critic Reinforcement Learning to
provide strong solutions. Through extensive experiments, we demonstrate GNOSIS outperforms various state-of-the-art algorithms for various network topologies while balancing storage usage and image access latency. Our approach aims to facilitate highly responsive edge application deployments critical for NextGen services.

A preliminary version of this paper appears as a part of our previous research published in the proceedings of the 2023 IEEE International Conference on Cloud Computing~\cite{10255001}.
In that conference paper, we initially proposed modeling the edge image placement problem as a minimum vertex cover and introduced the GNOSIS approach. We presented preliminary results that evaluated GNOSIS against a Greedy algorithm on various network topologies.
In this journal article, we provide a more comprehensive treatment of the problem through an extended discussion of the motivation and challenges of deploying microservices at the edge. Additionally, we significantly contribute expanded experimental results, evaluating GNOSIS against various state-of-the-art solutions, specifically Approximation, Greedy and Genetic. We empirically analyze and evaluate GNOSIS across different network topology variations, incorporating detailed performance metrics. This extended evaluation aims to further validate the effectiveness of our learning-based solution for real-world edge infrastructures.

The remaining of this paper is organized as follows. Sec.~\ref{related_work} contextualise our proposal with respect to the state-of-the-art scientific literature. Sec.~\ref{problem_formulation} formally defines the problem we are targeting, organizing the presentation around three main subtopics: Minimum Vertex Cover and Proactive Image Placement, Set Cover and Linear Optimization.
Sec.~\ref{mvc_for_pip} describes the different algorithmic approaches for targeting the minimum vertex problem problem. Sec.~\ref{proposed_approach} discusses our proposed approach. In Sec.~\ref{experimental_evaluation} are presented the experimental results we achieved. Finally, in Sec.~\ref{conclusion} we draw our concluding remarks. 

\section{Related Work}
\label{related_work}

The optimal placement of application images within the dynamic and heterogeneous Cloud-Edge Continuum stands as a pertinent issue \cite{8846920,kaur22}. Addressing this challenge involves leveraging various solution methodologies that are based on optimization techniques, such as Integer Programming \cite{7737202,8014365}, Markov Decision Processes~\cite{Puterman1994MarkovDP}, and stochastic optimization~\cite{neely2010stochastic}.
Several variations of the Integer Programming paradigm have been outlined in \cite{7737202,8014365,10.1007/s11761-017-0219-8}. In \cite{DBLP:journals/corr/abs-1802-00800}, authors explore the use of Integer Nonlinear Programming in the domain of Fog Computing, while Mixed Integer Linear Programming is suggested in \cite{8241089}. Another avenue for formulating the image placement problem involves implementing Constrained Optimization, as detailed in \cite{8814186}. Constrained Optimization comprises methods that identify optimal values for specific variables while adhering to specified constraints. 
Moreover according to \cite{Salaht2020AnOO}, Markov Decision Processes \cite{Puterman1994MarkovDP}, stochastic optimization \cite{6813406}, and general convex optimization are also some potential solutions to effectively formalize the image placement problem.

Additionally, the optimization problem's framing within the utilized metrics constitutes another important aspect observed in various research studies within the scientific literature. For instance, in \cite{Velasquez2017ServicePF,7987464,7996574}, the focus was on minimizing service latency in the image placement process. 
Low latency is considered integral in the context of delay-sensitive applications, thus steering the majority of the scientific efforts toward employing latency as the primary metric. Moreover, Reinforcement Learning techniques are explored in \cite{ROSSI2020161} for developing an extension of Kubernetes able to deploy and replicate delay-sensitive containerized services in a geographically distributed system. Furthermore, a solution called ICON is proposed in \cite{zavodovski2018icon}, where autonomous containers collaborate to find the best (in terms of latency) allocation of their services. 
Considering the nature of Edge and Cloud infrastructures, minimizing operational expenses emerges as a pivotal concern. Hence, it becomes evident that the cost associated with image placement is another measure that should be considered, as indicated in \cite{Arkian2017MISTFD}. Furthermore, resource utilization stands as another crucial metric, explored within optimization strategies, as evidenced in \cite{8014364} and \cite{10.1145/3057266}. Lastly, the congestion rate stands as an additional metric deserving exploration. Yu et al. \cite{8486269} explored the congestion ratio, investigating its potential as the minimum ratio between flow and link capacity to accommodate service placement.

Beyond technical methods, another consideration involves determining the approach for operational control implementation. Image placement processes can occur through centralized, distributed, or federated means. Within a centralized control logic, decisions consider all infrastructure resources and deployment processes. The majority of scientific efforts devoted to tackling the image placement problem, utilize some form of centralized control logic, as examined in \cite{8241101}.
In contrast, distributed control logic utilizes multiple nodes to establish an orchestration and management layer, facilitating image placement functionalities. This distributed layer formulates strategies and decisions based on locally harvested information, creating a robust control plane adaptable to the dynamic nature of Edge infrastructures. However, only a limited number of scientific endeavors have explored distributed control planes. For instance, Wang et al. \cite{Wang2018AFA} introduced a Fog Computing architecture reliant on coordinating various Fog nodes.

Previous efforts emphasize on reactive image placement strategies triggered after service requests in Edge computing environments. However, given the complex functionalities and demanding Quality of Service (QoS) requirements associated with contemporary services, this reactive paradigm seems insufficient. Our focus lies in implementing a proactive approach, a relatively unexplored area within the scientific community. Only a few exceptions, such as the work in \cite{9355698}, \cite{makris2024pro} and \cite{8538655}, have delved into proactive image placement. For instance, in \cite{9355698}, a Reinforcement Learning-based mechanism proactively deploys microservices on edge servers, considering the structure of the microservice graph application. Similarly, in \cite{8538655}, efforts aimed to establish a service placement and migration model leveraging mobility prediction in Mobile Edge Computing (MEC). More specifically, it proposes a virtual machine (VM) placement and migration decision model based on mobility prediction. An Integer Linear Programming (ILP) model is introduced to optimize the VM placement within candidate cloudlets. However, these studies primarily focus on specific solutions rather than thoroughly analyzing the potential effects of various algorithmic approaches or network topologies on proactive image placement.
 
In the last years, the success of graph neural networks has boosted the research on many combinatorial optimization problems \cite{huang2019review, khalil2017learning}, including the minimum vertex cover.
A novel hybrid method for solving the minimum weighted vertex cover is presented in \cite{langedal2022efficient}. This hybrid approach combines several strategies to create an effective heuristic including reduction rules, a GNN to classify nodes according to their likelability to be in the final solution and an exact solver to compute an initial vertex cover. Finally, the method uses an improved local search strategy to enhance the solution further. In \cite{abu2022learning}, a deep learning approach to solve the minimum vertex cover is presented. The paper introduces an obstruction-aware training strategy for the deep learning model. This strategy involves training the model to predict the likelihood of a vertex being in the MVC solution, given the presence of obstructions in its neighborhood. Then, the trained model is used to guide the search process, by prioritizing the exploration of vertices that have a high probability of being in the MVC solution.

All the aforementioned solutions represent specific approaches tailored to particular scenarios, overlooking the potential impact of network topology on final results. 
In this work, we formulate the proactive placement of application images as a minimum vertex cover problem and delve into a range of algorithmic approaches, spanning from traditional Approximation and Greedy methods to more advanced ones like the Genetic algorithm. Our exploration extends beyond theoretical propositions, suggesting a Metaheuristic approach which combines Graph Neural Networks and actor-critic Reinforcement Learning. The evaluation encompasses diverse network topologies such as Erd\"o-R{\'e}nyi, Watts-Strogatz and Barab{\'a}si-Albert.
What distinguishes our research is our comprehensive evaluation, integrating numerous metrics to assess algorithmic efficiency. Finally, our assessment goes beyond traditional considerations such as latency and operational costs. It spans a multi-dimensional spectrum, considering factors such as object volume, available link bandwidth, transfer time constraints, placement cost function, and node storage capacity.

\section{Problem Formulation}
\label{problem_formulation}

In this section, we delve into various modeling approaches aimed at addressing the problem of proactive image placement (PIP) in the Cloud-Edge continuum. This problem involves strategically allocating application images across the network system to minimize application runtime and enhance the overall performance of data-intensive and interactive applications. We explore three approaches that have been used to perform this modelling: Minimum Vertex Cover and Proactive Image Placement, Set Cover, and Linear Optimization.

\subsection{Minimum Vertex Cover and Proactive Image Placement}
The proactive image placement problem can be effectively modeled as a minimum vertex cover problem (MVC) in graph theory~\cite{karp1972reducibility,karakostas2009better,dinur2005hardness}, these problems are NP-complete. In this approach, the network system is represented as a graph, where devices are denoted as nodes and communication links between them as edges. The main objective is to identify a minimum vertex cover, which refers to selecting the minimum number of nodes needed to cover all edges in the graph.
The network is modelled as a graph $G = (V,E)$. Each $v \in V$ is a network node that can directly hold or obtain at 1-hop distance the image of the target application. 
Each $e \in E$ is an edge that represents a link between two nodes $(s,~d)~|~s \in V, d \in V$, where $s$ is the source node and $d$ is the destination node. 
Each edge $e$ has a given available bandwidth, which can be used to determine its weight. A valid solution $S$ should contain all nodes $v$ that request the image. The optimal solution, denoted $S \subseteq V$, comprises a list of nodes where the image needs to be placed, with the aim of minimizing the total transfer delay and the number of replicas.
By leveraging the minimum vertex cover formulation, it is possible to strategically place application images on the selected nodes, ensuring that all communication links are covered. This approach is particularly suitable for Edge computing environments, where computations are performed closer to the end-users, resulting in reduced latency and improved bandwidth utilization. Furthermore, proactive image placement based on the minimum vertex cover can adapt to the dynamic nature of Edge networks, allowing for efficient scaling and resource allocation.

\subsection{Set Cover}

In addition to the minimum vertex cover formulation, an alternative modeling approach for the proactive image placement problem is the set cover problem. 
The set cover problem involves selecting subsets from a given collection to cover a set of elements. In the context of proactive image placement, devices in the network represent the elements that need to be covered by selecting a subset of nodes to host the application images. 
The main constraint here is that all nodes $v \in V$ that request the image must have a link to at least one node that holds the image and is able to transfer it to $v$ within a given amount of time $T$, or be the one that hosts it.
A mathematical representation of the main constraint could be the following:
\begin{equation}\label{eq:1}
\frac{L_{image}}{W_{(s,~d)}} < T : \exists s \in DP_i,~\forall d \in V
\end{equation}
Where the value $L_{image}$ represents the byte length of the target image, and the value $W_{(s,~d)}$ represents the bandwidth of the link from source $s$ to destination $d$ node. The value of $T$ represents the time threshold under which the image must be completely transferred from $s$ to $d$. 
Every solution $DP_i$ is defined as a subset of nodes ($DP_i \subseteq V$) which will be used as source nodes $s$. 
A secondary constraint would be the minimization of the cost function associated with all nodes and edges involved in the solution.
For the solution $i$ which is defined as:
\begin{IEEEeqnarray}{lCr}
  DP_i = [Node_0,~Node_1,~\dots,~Node_s] \text{  and  } \\
  TC_i = \sum_{s \in DP_i}{F_{cost}(L_{image},R_s)}   
\end{IEEEeqnarray} 
Where the cost function is described by the function $F_{cost}(L_{image},R_s)$ having $R_s$ as the cost of disk space in node $s$. 
By formulating the problem as a set cover, we can explore different strategies to determine the optimal subset of nodes that covers all devices in the network. This alternative perspective allows to leverage existing algorithms and heuristics developed for the set cover problem. The set cover approach offers flexibility in terms of optimization techniques, scalability, and adaptability to different network topologies and constraints.

\subsection{Linear Optimization}
\label{opt_problem_section}

Linear optimization techniques, such as integer linear programming, provide yet another modeling approach for addressing the problem of proactive image placement.
Linear optimization allows us to formulate the problem as a mathematical program with defined objectives and constraints, which can be solved using optimization algorithms. By leveraging linear programming, we can obtain optimal or near-optimal solutions for the proactive image placement problem.
The formulation involves defining decision variables, objective functions, and constraints that capture the requirements and constraints of the proactive image placement problem. The objective function can be designed to minimize the overall application runtime, considering factors such as network latency, bandwidth utilization, and resource constraints. The constraints can include limitations on the number of images that can be placed on each device, communication delays, and energy consumption.
While linear optimization techniques offer the potential for obtaining optimal solutions, it is important to consider the computational complexity and scalability challenges associated with solving large-scale instances of the problem. Advanced optimization algorithms and techniques, such as decomposition methods or approximation algorithms, may be required to address these challenges effectively.
%
%
In the context of our problem, a set of binary variables, donated as $A_v$ where $A_v \in \{0,1\}$ and $v \in V$, needs to be defined. Variables are assigned the value of $1$ when the referenced node $v$ can provide the image, indicating its role as a source node, and $0$ otherwise.
Such variables are used to formulate basic constraint, as follows:
\begin{equation}
\sum_{v}(A_{v} \cdot  \frac{L_{image}}{W_{(v,~d)}}) > 0 ~\mid ~\forall v \in V, ~\forall d \in D
\end{equation}
with $L_{image}$ is still the byte length of the image, $W_{(v,~d)}$ the bandwidth available for the link between source node $v$ and node $d$ belonging to the set of destination nodes $D$.
This function describes the constraint that forces the optimizer to cover the needs of each destination node by having at least one valid link for each destination node.
To formulate the objective function, we need to explicitly define two costs; the total transfer time and the total occupied disk space. 
The definition of a constraint that forces transfer delay limitations is impossible in classical optimization problem formulation so this constraint need to be relaxed into a ``best effort'' one. 
This means that transfer time will be integrated into the objective function, aiming at the minimization of the total transfer time. 
This will, in theory, limit the time threshold violations as much as possible throughout the network.
Taking all these factors and assumptions into consideration the cost function takes the following form:
\begin{IEEEeqnarray}{rCl}
min(\sum_{v}(A_{v} \cdot L_{image}) + \sum_{v}(A_{v} \cdot  \frac{L_{image}}{W_{(v,~d)}})) \nonumber \\ 
A_{v} \in \{0,1\}, \forall v \in V, \forall d \in D
\end{IEEEeqnarray}
An apparent issue with this formulation arises from the imbalance between the two main factors of the cost function since they are on different scales. 
As a result, even a slight adjustment to the first factor would exert a considerably more substantial influence on the function compared to an identical change in the second factor.
%
On the other hand, the second factor, which is the the transfer time, is expressed in milliseconds or seconds. In order to tackle this issue and mitigate its effects, the $L_{image}$ value from the first scale is removed, changing it from the total disk space to the total number of replicas in the network. This change balances the two factors better, enabling us to incorporate the number of replicas as a weight on the transfer time. Consequently, the transfer time becomes the predominant factor in the function.
The following represents the final form of the objective function contributing to the resolution of the MVC problem:
\begin{IEEEeqnarray}{rCl}
\label{final_objective_function}
min(\sum_{v}A_{v} + \sum_{v}(A_{v} \cdot  \frac{L_{image}}{W_{(v,~d)}})) 
\nonumber \\ 
A_{v} \in \{0,1\}, \forall v \in V, \forall d \in D
\end{IEEEeqnarray}

\section{Algorithmic Approaches to Minimum Vertex Cover Problem}
\label{mvc_for_pip}

The Minimum Vertex Cover problem is an extensively researched combinatorial optimization problem with various applications across domains such as network security, scheduling, feature selection, and more.
It is also closely related to other graph problems like the independent set, dominating set, and clique problems. As a result, numerous approximated algorithms have been proposed over the years to construct the vertex cover $S$ in different ways. 
In what follows, different approaches to solve the problem are presented. The network is treated as a graph where each node can \emph{(i)} host an application image, \emph{(ii)} require an application image or \emph{(iii)} do both or none of the above.


\subsection{Approximation}
\label{approximation_algo}

An approximation algorithm is an algorithm that offers a solution to an optimization problem, ensuring that it is reasonably close to the optimal solution. In the context of the MVC problem, there exist approximation algorithms that take as input an undirected graph $G$ and returns a vertex cover whose size is guaranteed to be no more than twice the size of an optimal vertex cover~\cite{cormen2009introduction}.
The algorithm takes advantage of the fact that vertices with higher out degrees are more likely to be included in the minimum vertex cover.
The algorithm starts with an empty vertex cover and repeatedly adds vertices to it until all edges in the graph are covered. Specifically, the algorithm repeatedly selects an arbitrary edge in the graph, adds both endpoints (nodes) of the edge to the vertex cover, and then the edge and its neighboring edges are removed from the graph. By doing so, the algorithm guarantees that all edges connected to these vertices are covered. 
Since the nodes added to the solution already cover all edges in the removed set, there is no need to further consider them. This process is repeated until every edge in the graph has been removed, guaranteeing that the solution will cover all edges in the graph. 
The algorithm is relatively simple and fast compared to other algorithm options due to its structure. This makes the algorithm a popular choice for the MVC problem. However, the algorithm does not always provide an exact solution, except for particular cases, which can be disadvantageous depending on the application.

\subsection{Greedy}
\label{greedy_algo}

The greedy approach to algorithm design is a very efficient method that can be used to solve various optimization problems efficiently. Its approach is to always choose the option that seems to be the best at the current moment, aiming to maximize the gain based on local conditions, assuming that this will lead to a globally optimal solution~\cite{cormen2009introduction}.
When applied to the minimum vertex cover problem, the Greedy algorithm starts by initializing an empty set of vertices $S$.
It then selects an arbitrary edge $e \in E$ connecting nodes $v_i, v_j \in V$ in the graph $G=(V,E)$, it adds $v_i, v_j$ to $S$, and finally removes all the edges from the graph that are covered by the vertices in $S$. 

\begin{IEEEeqnarray}{lCr}\label{eq:sets}
    e = (v_i, v_j) \in E,~v_i, v_j \in V \\
    S' = S \cup \{v_i, v_j\} \\
    E' = E \setminus \{(u, w) \in E \mid u, v \in S \}
\end{IEEEeqnarray}

If the graph is empty, the algorithm returns $S'$ as the minimum vertex cover. If not, the process starts again by selecting another arbitrary edge $e$.
At each step, the algorithm chooses the edge that has the smallest number of uncovered vertices and includes both endpoints in the vertex cover set. Therefore, this approach ensures that the set of vertices will cover as many edges as possible.
The Greedy algorithm does not always produce an optimal solution, as in some cases, the algorithm may get stuck in a local optimum and fail to find the global optimum. However, in the case of the minimum vertex cover problem, it is guaranteed to produce a solution that is at most twice the size of the optimal solution. This bound is known as the 2-approximation guarantee.

\subsection{Genetic}
\label{genetic_algo}

Genetic Algorithms (GAs) harness evolutionary principles such as natural selection, genetic variation, and the survival of the fittest observed in biological organisms to create heuristic search algorithms~\cite{holland1992adaptation}. Inspired by Darwinian evolution, GAs employ concepts like crossover, mutation, and natural selection to intelligently explore a defined search space and tackle complex problems~\cite{kotecha2003hybrid}.
In scenarios characterized by vast exploration spaces and the availability of fitness evaluations for solutions, GAs prove highly effective. Take, for instance, the Minimum Vertex Cover (MVC) problem, where the solution space is expansive, and fitness evaluations can gauge the minimality of the cover. Hence, GAs present a viable approach for addressing the MVC problem.
The evolutionary process begins by initializing a population of chromosomes, each representing a potential solution to the MVC problem. These chromosomes are generated either randomly or using heuristic methods. Parent chromosomes are then chosen based on their fitness values~\cite{goldberg1989genetic}, which reflect how effectively their corresponding sets of vertices cover edges in the graph. 
Through stochastic selection and modification employing crossover and mutation operators, new populations emerge, inheriting advantageous traits from the fittest individuals of the preceding generation. This iterative cycle persists until an optimal solution is attained or a predefined termination condition is met.
The optimization mechanism of a genetic algorithm relies on pivotal elements such as the fitness function, encoding scheme, crossover, and mutation. Together, these components drive the generation of increasingly fit solutions. The output of a genetic algorithm is the chromosome with the highest fitness value, representing a minimum vertex cover that efficiently addresses the MVC problem.

The genetic algorithm was configured with the following parameters: a) Population Size: 100 individuals per generation (balance between exploration and computational efficiency), b) Generations: 150 (sufficient iterations for the algorithm to converge to a near-optimal solution), c) Selection Method: Roulette wheel selection was utilized to choose individuals from the current population as parents for generating offspring (a common method that ensures individuals with higher fitness have a higher chance of being selected), d) Crossover Method: Order Crossover (OX) was applied to combine genetic material from selected parents (ensures the preservation of the relative order of genes from parent chromosomes) and e) Mutation Rate: A mutation rate of 0.1 was set, indicating a 10\% chance of mutation for each gene in an individual solution during each generation of the genetic algorithm (strikes a balance between exploration and exploitation).


\section{The GNOSIS Approach}
\label{proposed_approach}

The GNOSIS approach for addressing the MVC problem relies on the use of Graph Neural Networks \& Deep Reinforcement Learning in order to perform combinatorial optimization.

\subsection{Graph Neural Networks}
Graph Neural Networks are a type of neural network that operate on graph-structured data, allowing for the learning of node and graph-level representations. They can be used for a variety of tasks, including node classification, graph classification, link prediction and accurate resource usage prediction~\cite{theodoropoulos2023graph}. Here we will focus on explaining GNNs for graph-level representations.

Let us consider an undirected graph $G = (V, E)$, where $V$ is the set of nodes and $E$ is the set of edges. The graph can be represented by an adjacency matrix $A \in \mathbb{R}^{n \times n}$, where $n$ is the number of nodes. The elements of $A$ denote the presence or absence of edges between nodes, such that $A_{ij} = 1$ if there is an edge between nodes $i$ and $j$, and $A_{ij} = 0$ otherwise. In addition, we assume that each node $i$ has an associated feature vector $x_i \in \mathbb{R}^d$, where $d$ is the number of features.

The goal of GNNs is to learn a function that maps the graph $G$ and its associated node features ${x_i}_{i \in V}$ to a vector representation of the entire graph. 
%
%
Tipically, GNNs achieve this by organizing a sequence of neural network layers, where some layers aggregate information from neighboring nodes and update node representations, consequently.
Following this approach, at each layer $k$, the node representations are updated according to the following formula:

\begin{equation}
h_i^{(k)} = \sigma \left( \sum_{j \in \mathcal{N}(i)} \frac{1}{c_{ij}} W^{(k)} h_j^{(k-1)} + b^{(k)} \right)
\end{equation}
where $h_i^{(k)}$ is the updated representation of node $i$ at layer $k$, $\mathcal{N}(i)$ is the set of neighboring nodes of node $i$, $c_{ij}$ is a normalization constant, $W^{(k)}$ and $b^{(k)}$ are the trainable weight matrix and bias term at layer $k$, and $\sigma$ is an activation function such as the rectified linear unit (ReLU).
After $K$ layers, the final representation of the graph can be obtained by applying a readout function to the node representations, such as summation or max-pooling:

\begin{equation}
h_G = \rho \left( \sum_{i \in V} h_i^{(K)} \right)
\end{equation}
where $h_G$ is the final representation of the graph and $\rho$ is the readout function.

The GNN is trained by minimizing a loss function that depends on the final graph representation and the true label or target associated with the graph, using backpropagation and stochastic gradient descent. GNNs allow for the learning of graph-level representations by recursively aggregating information from neighboring nodes through multiple layers of neural network operations. This approach can be used for a variety of tasks, including combinatorial optimization problems.

\subsection{The Actor-Critic Algorithm}
Actor-critic is a class of reinforcement learning algorithm that combines the advantages of both policy-based and value-based methods. It consists of two neural networks: an actor network, which learns the policy, and a critic network, which estimates the value of the policy.

When leveraging temporal difference (TD) learning approaches, the critic network learns to estimate the expected cumulative reward by iteratively updating its value function based on the observed rewards. Specifically, at each time step $t$, the critic updates its value estimate $V_\phi(s_t)$ using the TD error:

\begin{equation}
\delta_t = R_t + \gamma V_{\phi}(s_{t+1}) - V_{\phi}(s_t)
\end{equation}
where $R_t$ is the observed reward at time step $t$, $\gamma$ is the discount factor, and $s_t$ and $s_{t+1}$ are the current and next states, respectively. The TD error represents the difference between the predicted reward and the actual reward, and is used to update the critic's value estimate:

\begin{equation}
V_{\phi}(s_t) \leftarrow V_{\phi}(s_t) + \epsilon\delta_t 
\end{equation}
where $\epsilon$ is the corresponding learning rate.

In the actor-critic algorithm with TD learning, the actor network learns the policy by maximizing the expected cumulative reward, which is estimated using the critic's value function. The policy update is based on the advantage function $A_t$, which represents the advantage of taking action $a_t$ in state $s_t$ compared to following the current policy. The advantage function is defined as:

\begin{equation}
A_t = \delta_t + \gamma V_{\phi}(s_{t+1}) - V_{\phi}(s_t)
\end{equation}
Finally, the policy update is then given by:
\begin{equation}
\theta \leftarrow \theta + \beta\nabla_{\theta} \log \pi_{\theta}(a_t|s_t) A_t
\end{equation}
where $\theta$ are the parameters of the actor network, $\beta$ is the learning rate, $\log \pi_\theta(a_t | s_t)$ is the log-probability of taking action $a_t$ in state $s_t$ according to the policy, and $\nabla_\theta$ is the gradient operator.

\subsection{Graph Neural Networks \& Deep Reinforcement Learning for Combinatorial Optimization}

Actor-critic algorithms have shown promising results in solving combinatorial optimization problems. By combining actor-critic with GNNs, we can leverage the power of GNNs to represent graph structures and use actor-critic RL to provide solutions.
GNNs provide an efficient way to represent graph structures, capturing complex dependencies and relationships among vertices and edges, which is particularly advantageous in edge computing environments where efficient representation is essential for optimization tasks. When combined with a DRL approach, GNNs can offer adaptability and flexibility in solving optimization problems (e.g. MVC), enabling the system to dynamically adjust strategies in response to evolving network conditions, computational requirements and latency constrains \cite{munikoti2023challenges}.

One may represent a combinatorial optimization problem as a graph $G=(V,E)$, where $V$ is the set of vertices and $E$ is the set of edges. Each vertex represents a decision variable, and each edge represents a constraint between two decision variables. Let $x\in{0,1}^n$ be a binary decision vector, where $n=|V|$, such that $x_i=1$ if vertex $i$ is selected and $x_i=0$ otherwise. The goal is to find an optimal decision vector $x^*$ that maximizes an objective function $f(x)$, subject to the constraints encoded in the graph.

To apply actor-critic algorithms with GNNs, one has to define a policy function $\pi_{\theta}(a|s)$ that takes as input a state representation $s$ and outputs a probability distribution over the action space $a$. We parameterize the policy function with $\theta$. The state representation $s$ is obtained by feeding the graph structure and the current decision vector $x$ through a GNN. The actor and critic are updated using the policy gradient and the TD learning algorithms, respectively. The policy gradient updates the policy parameters $\theta$ using the gradient of the expected reward with respect to $\theta$. The TD learning updates the critic parameters $\phi$ by minimizing the mean squared error between the estimated and actual values.

The actor-critic with GNNs algorithm for combinatorial optimization is described in Algorithm \ref{Alg:Algo1}.
\begin{algorithm}
\caption{Actor-critic with GNNs algorithm for combinatorial optimization}  
\label{Alg:Algo1}
\begin{algorithmic}
\STATE \textbf{Begin}\\
{1.}Initialize the policy and critic parameters $\theta$ and $\phi$.\\
{2.}For each episode do:\\
{3.}\hspace*{3mm} Initialize the decision vector $x$ randomly.\\
{4.}\hspace*{3mm} Feed the graph structure and $x$ through the GNN to obtain the state representation $s$.\\
{5.}\hspace*{3mm} Sample an action $a$ from the policy distribution $\pi_{\theta}(a|s)$.\\
{6.}\hspace*{3.5mm}Obtain the next state $s'$ and reward $r$ by applying the action $a$ to $x$.\\
{7.}\hspace*{3.5mm}Update the critic parameters $\phi$ using TD learning.\\
{8.}\hspace*{3.5mm}Compute the policy gradient and update the policy parameters $\theta$.\\
{9.}\hspace*{3mm} Repeat steps 4-8 until a stopping criterion is met.\\
{10.}\hspace*{2.5mm}Obtain the next state $s'$ and reward $r$ by applying the action $a$ to $x$.\\
{11.}End For.\\
{12.}Return the decision vector $x^*$ that maximizes $f(x)$.\\
\STATE \textbf{End}

\end{algorithmic}
\end{algorithm}
More specifically, in the case of solving the MVC problem, GNNs and the actor-critic algorithm are leveraged in the following manner. As stated before, GNNs are employed for state representation, where the state $s$ passed through the GNN encompasses both the current vertex cover and indicators for uncovered edges. This implementation design enables the GNN to prioritize vertices crucial for covering the most uncovered edges, thereby aiding in minimizing the vertex cover set size. Subsequently, action selection, governed by the policy $\pi_\theta(a|s)$, strategically chooses vertices likely to enhance cover efficiency, with a particular emphasis on those linking to multiple uncovered edges. Moreover, the reward $r$ is carefully crafted to balance between the size of the vertex cover and the necessity to cover all edges. Immediate rewards are given for actions that lead to a significant reduction in uncovered edges, incentivizing the selection of highly connected vertices not yet in the cover. Finally, the actor-critic algorithm dynamically updates the state representation $s$ with each action, allowing the GNN to adjust its focus on the remaining uncovered edges. This iterative refinement is key to adapting the vertex cover in response to the evolving graph coverage.

The code for the proposed approach is open source and available for access
\footnote{\url{https://github.com/f-coda/GNOSIS}}.

\section{Experimental Evaluation}
\label{experimental_evaluation}

This section details the experimental evaluation of the GNOSIS approach in relation to various performance indicators. The examined algorithms and GNOSIS are subjected to a rigorous investigation through extensive experiments in the minimum set cover problem, revealing valuable insights into their properties and impacts on proactive image placement.
In the following, we describe the simulation methodology (subsection \ref{sim_methodo}) and analyze the experimental results (\ref{sim_results}).  

\subsection{Simulation methodology}
\label{sim_methodo}

To simulate various network topologies, as well as image placement and image transfers between nodes, Python scripts (version 3.6.9) were developed and utilized.
More specifically, the NetworkX Python package \cite{hagberg2008exploring} was leveraged to generate diverse network topologies with varying parameters. Furthermore, NetworkX constructs objects that facilitate the storage and manipulation of node and edge labels and attributes, including total capacity, image replication, and network connectivity utilization.

Each network connection, represented as an Edge, is characterized by two attributes: available bandwidth and usage. These attributes enable the derivation of various secondary values, such as transfer time and the percentage of bandwidth usage. On the other hand, each node is assigned a unique serial number/ID and has an attribute with a Boolean value that denotes whether it holds a replica of the image (1) or not (0). Additionally, several parameters, such as image size and the maximum available bandwidth for Ethernet and WiFi connections, were established for all experiments, irrespective of the networks involved.

When evaluating content distribution systems by means of simulation, it is of utmost importance to correctly mimic the bandwidth dynamics behaviour of the underlying network. In the simulations carried out in this study, bandwidth allocation was accomplished using the Max-min fairness algorithm. Max-min fairness \cite{bertsekas2021data} aims to optimize the allocation of bandwidth to flows with a minimum share, ensuring that no flow can increase its rate at the expense of another flow with a lower rate. This is achieved by initially setting all flow rates to zero and then increasing each rate equally until the link's capacity is reached. The simulated network is modeled as an undirected graph, and Max-min fairness can provide a reliable estimate of the network's actual behavior. Compared to the equal sharing policy, Max-min fairness achieves higher average throughput and more efficient resource utilization.  On the other hand, while maximum throughput resource management produces higher average throughput than Max-min fairness, it may lead to the starvation of costly flows.
As a total bandwidth capacity, the maximum bandwidth of the node's virtual network adapter is considered, which is randomly set to either Ethernet (100MBps) or WiFi (25MBps). In practical Edge networks, the majority of communication occurs over WiFi due to the widespread use of smart devices and IoT equipment. Therefore, the WiFi network adapters were arbitrary set to a 75\% ratio while Ethernet adapters were assigned to the rest 25\% of nodes.

The performance of the examined algorithms is assessed on three different network graph topologies, each of which is simulated at varying sizes ranging from $64$ to $512$ vertices, including $V = [64, 128, 256, 512]$. 
The variation in the number of edges in each graph is largely influenced by the unique features, input parameters, and connectivity properties inherent to each topology, which in turn depend on the varying number of vertices present.

The network topologies utilized in this work are as follows:

\textbf{Erd\"o–R{\'e}nyi random network}. Paul Erd{\"o}s and Alfr{\'e}d R{\'e}nyi first introduced the concept of random graphs, discovering that probabilistic techniques were effective in solving graph theory problems \cite{gilbert1959random,erdos1959random}.
During the introduction of these network types in the 1950s, computing power was limited, leading to a primary emphasis on modeling relatively small ``ordered'' or ``regular'' networks, which are infrequent in real-world scenarios \cite{albert2002statistical}. 
An equivalent and alternative definition of a random graph is the binomial model, in which the $G(N,p)$ model starts with $N$ nodes and connects each distinct node pair with probability $p$.
This work considers three different variants of the Erd\"o–R{\'e}nyi model, which are based on the probability $p$ indicating an edge between any two nodes: $p=0.2$, $p=0.5$ and $p=0.7$. As expected, the structure of the graph generated by each probability, varies considerably; as $p$ increases, the number of edges also increases.

The uncorrelated Erd\"o–R{\'e}nyi random graph model assumes that every pair of vertices in a graph has equal and independent probabilities, thus treating the network as a collection of equivalent units. However, real networks are inherently correlated systems, and their topology often deviates from that of the uncorrelated random graph model. Consequently, more sophisticated graph models have been developed, with a particular focus on ``real-world'' networks such as the Internet and the World-Wide Web. To understand the general properties of such networks, two classes of models have emerged: ``small-world'' and ``scale-free''. Small-world networks aim to capture the clustering observed in real graphs and exhibit heterogeneity in that the pattern of connections between nodes is relatively localized. On the other hand, scale-free networks exhibit heterogeneity in the ``degree'' of nodes (i.e., the number of connections a node has to other nodes) and reproduce the power law degree distribution present in many real networks.

\textbf{Watts–Strogatz small-world network}. The archetypical small-world network was proposed by Watts and Strogatz \cite{watts1998collective}. The algorithm behind the model begins by constructing an undirected ring lattice network, consisting of a ring of nodes with edges evenly distributed between its $k_{L}$ nearest left and right neighbors, where $k_{L}$ represents the degree of each node in the initial lattice. Then, a random rewiring process is applied, where each edge has a probability $p$ of being rewired. The algorithm only rewires one end of each edge and traverses edges in a way that ensures that each node loses at most half of its edges. 
It's worth noting that edges are only replaced, not added or removed, so the total number of edges and the mean degree remains constant. 
By varying $p$, it can be shown that only a few rewires are needed to produce a low average path length while maintaining a high clustering coefficient. In fact, for $p=0$, the small-world model preserves a regular graph, while $p=1$ generates a random graph, which differs only slightly from the uncorrelated random graph. For intermediate values of $p$, Watts-Strogatz produces a small-world network, which captures the high clustering properties of regular graphs and the small characteristic path length of random graph models. Hence, in his work we focused on a single rewiring probability of $p=0.5$ for three distinct degree values: $k_{L}=2$, $k_{L}=4$ and $k_{L}=7$. As the degree of a node represents the number of edges connecting it to other nodes within the graph, the resulting graph structure differs significantly for each degree value; as $k_{L}$ increases, the number of edges also increases.

\textbf{Barab{\'a}si-Albert scale-free network}. Barab{\'a}si and Albert \cite{barabasi1999emergence} introduced the scale-free network that is capable of reproducing networks with ``hubs'', where a small number of nodes have significantly more connections than the average (scale-free property), resulting in a highly inhomogeneous degree distribution.
This model is widely used due to the fact that many real-world networks exhibit degree distributions similar to the Barab{\'a}si-Albert model. To create a Barab{\'a}si-Albert scale-free network of size $N$, the algorithm starts with a small number of nodes $m_o$, and sequentially introduces the remaining $N - m_o$ nodes into the network, where each node connects to/from $m \leq m_o$ existing nodes. It is standard practice to choose $m_o = m$. The maximum mean degree of the network is determined by the initial network size $m_o$, as selecting $m > m_o$ would result in the first newly introduced node not being able to be assigned $m$ edges. Three different variants of the Barab{\'a}si and Albert model have been considered in this work:  $m=1$, $m=3$ and $m=8$. As expected the structure of the graph generated by each value $m$, varies significantly; as $m$ increases, the number of edges also increases.

The probability $P$ that a new node will be connected to existing node $i$ (referential attachment) depends on the degree $k_{i}$ of node $i$, as $P(k_{i}) = \frac{k_{i}}{\sum_j k_{i}}$. 
The combination of network growth with this preferential attachment is what leads to a power-law degree distribution.
In contrast to the small-world model, the degree distribution in scale-free graphs follows a power-law when
$N \rightarrow \infty$, which has been shown to be a combined effect of growth and the preferential attachment \cite{barabasi1999mean}. Therefore, in the infinite time or size limit, the scale-free model has no characteristic scale in the degree size.

\subsection{Simulation Results}
\label{sim_results}

This section showcases the simulation results for the different network topologies obtained by the examined algorithms. To evaluate the efficacy of each algorithm in addressing the image placement problem, three performance metrics are utilized. These metrics include:
\begin{itemize}
    \item execution time ($ExT$): total amount of time each algorithm requires to generate a solution
    \item cost function ($CF$): calculates the cost based on the number of image replicas placed on the network as well as the transfer delays, in order to share the image between all network nodes. The cost function is the same as the objective function (Function \ref{final_objective_function}), as described in \ref{opt_problem_section}.
    \item size of the vertex cover ($VCS$): size of vertices in it
\end{itemize}

As discussed in subsection \ref{sim_methodo}, several variants are considered for each network topology, based on the input parameters of the respective model. However, due to the considerable number of experiments and limited space, it is impractical to visually represent the results for all variations of each network topology. Therefore, only one representative variation is considered for each model: $p=0.2$ for Erd\"o–R{\'e}nyi, $k_{L} = 2$ for Watts–Strogatz, and $m=1$ for Barab{\'a}si-Albert. Nonetheless, the detailed results for the remaining variants of each model are provided in Table \ref{tab:detailed_resultsall}. The abbreviations BA, ER, and WS used in the table, correspond to Barab{\'a}si-Albert, Erd\"o–R{\'e}nyi and Watts–Strogatz, respectively.

\begin{figure*}[!h]
\centering
\subfloat[Erd\"o–R{\'e}nyi]{\includegraphics[width=2.3in]{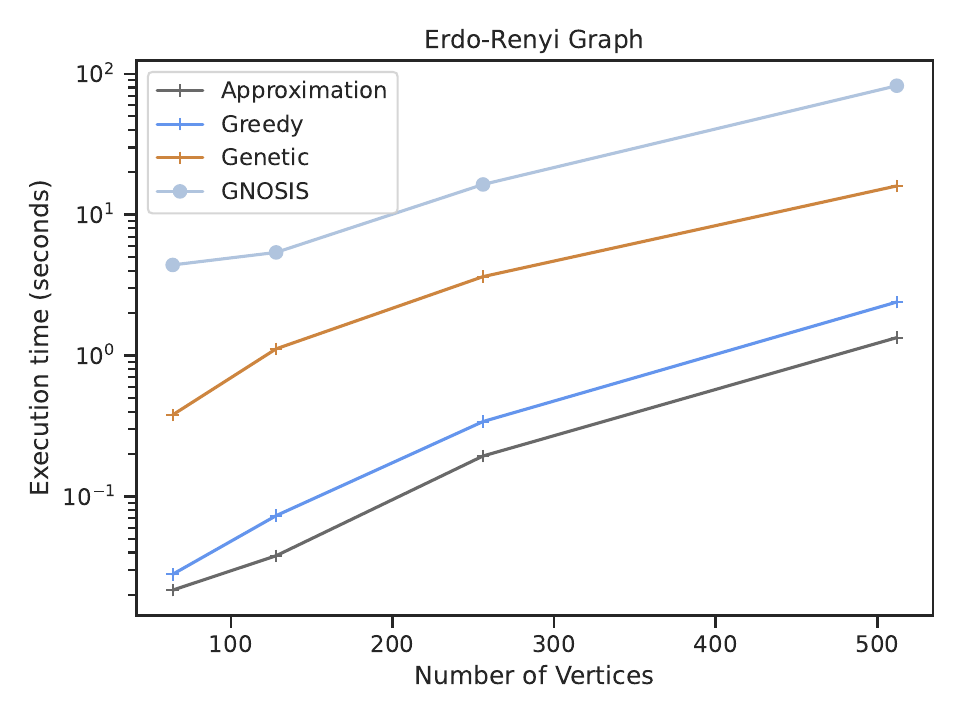}%
\label{ex_time_ER}}
\hfil
\subfloat[Watts–Strogatz]{\includegraphics[width=2.3in]{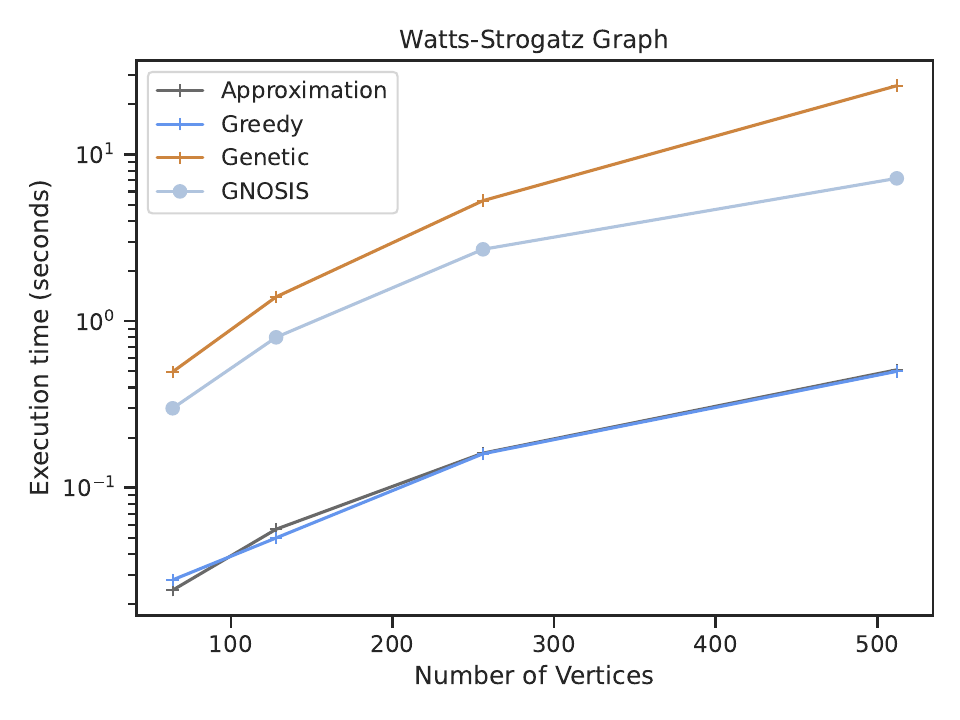}%
\label{ex_time_WS}}
\hfil
\subfloat[Barab{\'a}si-Albert]{\includegraphics[width=2.3in]{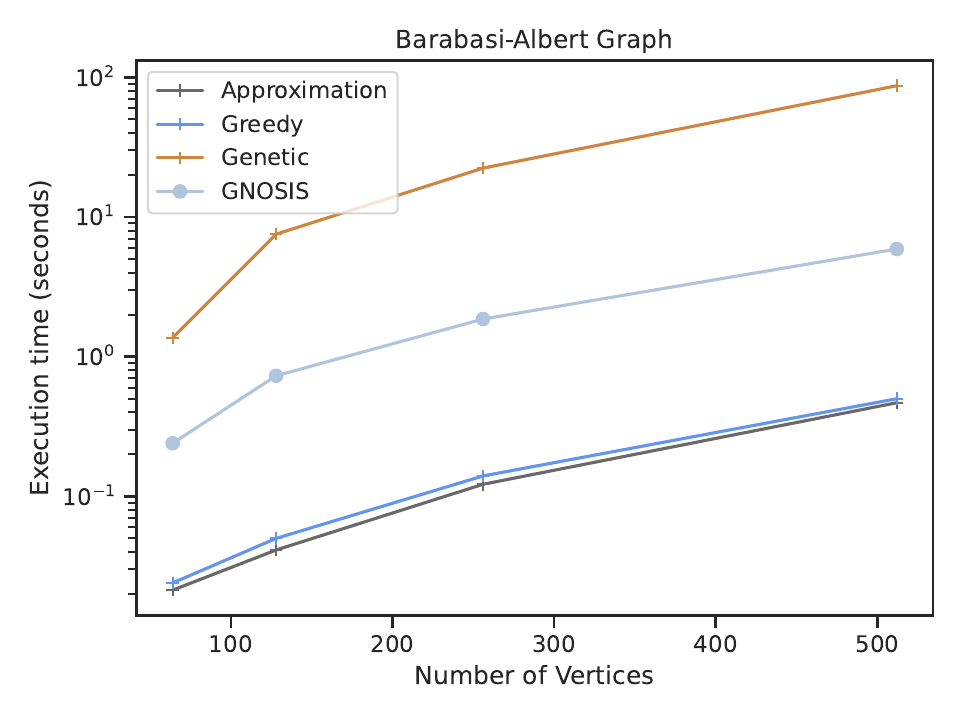}%
\label{ex_time_BA}}
\caption{Execution time of each algorithm for the different network topologies}
\label{execution_time_all}
\end{figure*}

\textit{Execution time analysis.} Figure \ref{execution_time_all} evaluates the execution time of each algorithm for the different network topologies. 
As the results indicate, GNOSIS has a notably longer execution time in Erd\"o–R{\'e}nyi graphs, particularly as the number of vertices grows.
Although the aforementioned graph topology results in longer execution times for both Greedy and Approximation algorithms, the difference is relatively small compared to other network topologies.
As an illustration, when considering $512$ vertices, the GNOSIS algorithm demonstrates an execution time of $82$ seconds for Erd\"o–R{\'e}nyi graphs and $5.9$ seconds (almost 14 times slower) for Barab{\'a}si-Albert graphs, whereas for the Greedy algorithm, the execution times are $2.4$ seconds and $0.5$ seconds (almost 5 times slower), respectively. 
Conversely, for the Barab{\'a}si-Albert and Watts-Strogatz topologies, the Genetic algorithm displays the longest execution time. In fact, for the Barab{\'a}si-Albert, the execution time is noticeably higher than that of the other algorithms.
The Approximation and Greedy algorithms display comparatively shorter execution times across all network topologies, with a significant difference when compared to the other algorithms.
All algorithms display a similar trend across the various network topologies, wherein the execution time increases linearly with an increase in the number of vertices. In general, as the number of vertices increases and more nodes are added to the vertex cover set, the execution time also increases as the generated graph becomes larger.

\begin{figure*}[!h]
\centering
\subfloat[Erd\"o–R{\'e}nyi]{\includegraphics[width=2.3in]{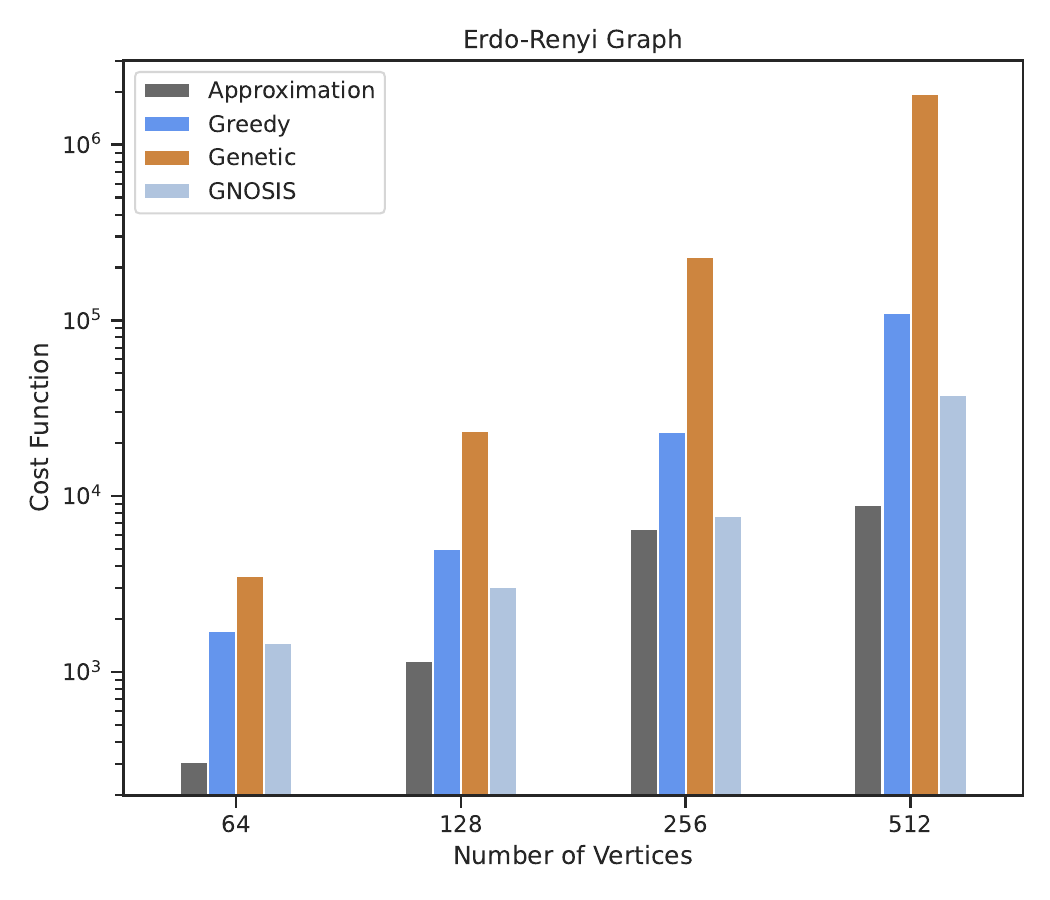}%
\label{cost_function_ER}}
\hfil
\subfloat[Watts–Strogatz]{\includegraphics[width=2.3in]{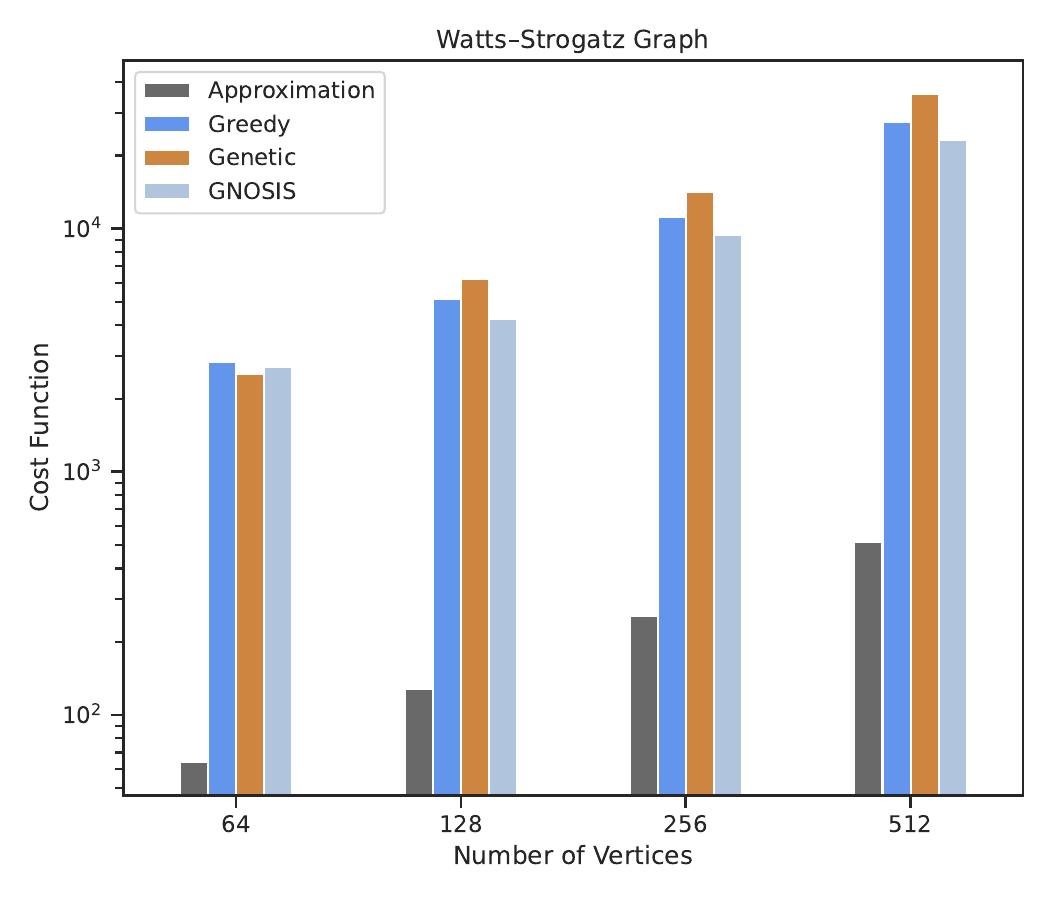}%
\label{cost_function_WS}}
\hfil
\subfloat[Barab{\'a}si-Albert]{\includegraphics[width=2.3in]{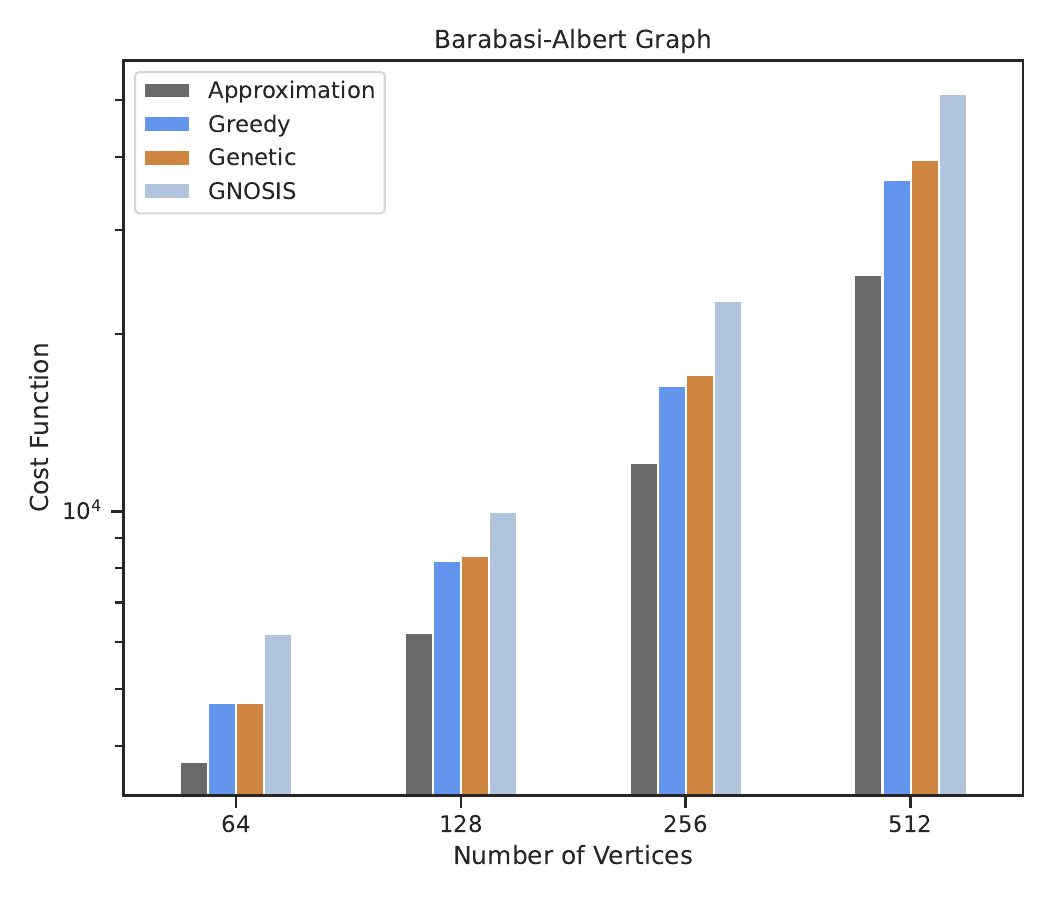}%
\label{cost_function_BA}}
\caption{Cost function of each algorithm for the different network topologies (the lower the better)}
\label{cost_function_all}
\end{figure*}

\textit{Cost function analysis.} Figure \ref{cost_function_all} evaluates the cost function of each algorithm for the different network topologies.
The results indicate that for Erd\"o–R{\'e}nyi graphs, the Genetic algorithm produces the highest cost function values, followed by the Greedy algorithm. Conversely for Barab{\'a}si-Albert graphs, the GNOSIS algorithm produces the highest cost function values.
To provide an example, when considering $512$ vertices, the Genetic algorithm yields cost function values of $1942119.08$, $39449.04$ and $35845.13$ for Erd\"o–R{\'e}nyi, Barab{\'a}si-Albert and Watts-Strogatz graphs, respectively. 
Notably, the cost function value for Erd\"o–R{\'e}nyi graphs is much higher than those for Barab{\'a}si-Albert and Watts-Strogatz graphs, with ratios of approximately $49.22$ and $54.19$, respectively.
Furthermore, in comparison to the other algorithms, the Genetic algorithm results in a significantly higher cost function values for Erd\"o–R{\'e}nyi graphs. Specifically, when considering $512$ vertices, the cost function values obtained by GNOSIS, Greedy, and Approximation algorithms are $37865.43$, $110230.25$, and $8951.86$, respectively.
For all algorithms, as the number of vertices increases, Watts-Strogatz graphs demonstrate comparatively lower cost function values when compared to the other network topologies. In addition, as the number of vertices increases, there is no significant difference in the cost function values between the examined algorithms for Watts-Strogatz graphs, except in the case of Approximation algorithm. 
In fact, the Approximation algorithm yields the lowest cost function values across all network topologies, with the discrepancy being particularly pronounced for Watts-Strogatz graphs.
It is worth noting that in both Erd\"o–R{\'e}nyi and Watts-Strogatz graphs, the GNOSIS algorithm demonstrates lower cost function values in comparison to the Greedy and Genetic algorithms.
The reduction in the cost function achieved by the GNOSIS algorithm is especially noteworthy for of Erd\"o–R{\'e}nyi graphs, as the decrease is nearly half with an increase in the number of vertices when compared to the Genetic algorithm.

\begin{figure*}[!h]
\centering
\subfloat[Erd\"o–R{\'e}nyi]{\includegraphics[width=2.3in]{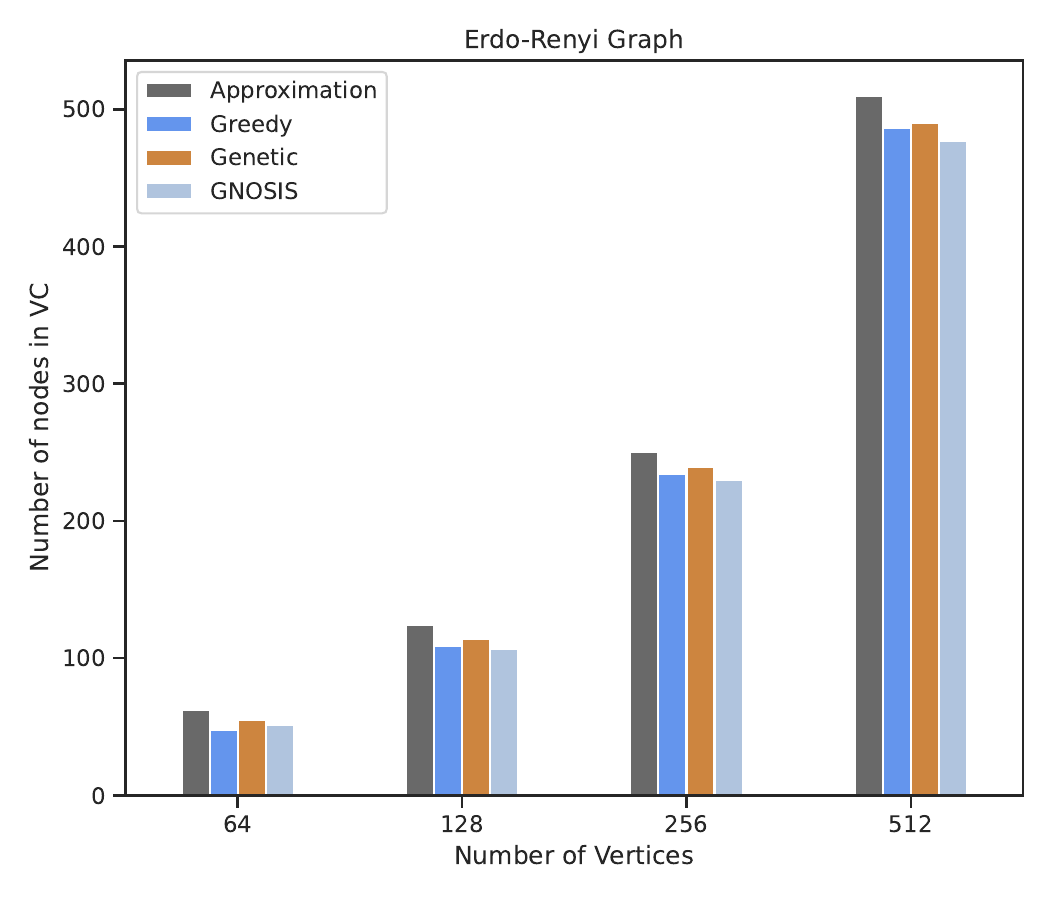}%
\label{VC_ER}}
\hfil
\subfloat[Watts–Strogatz]{\includegraphics[width=2.3in]{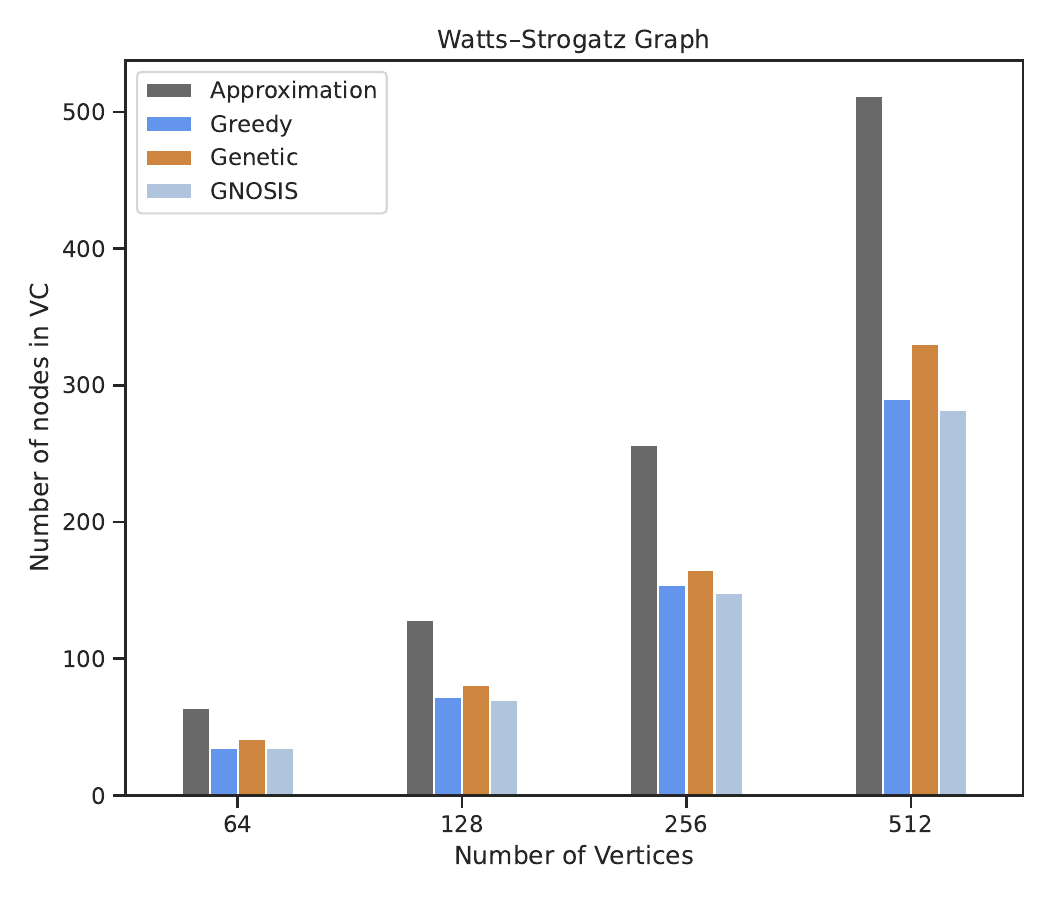}%
\label{VC_WS}}
\hfil
\subfloat[Barab{\'a}si-Albert]{\includegraphics[width=2.3in]{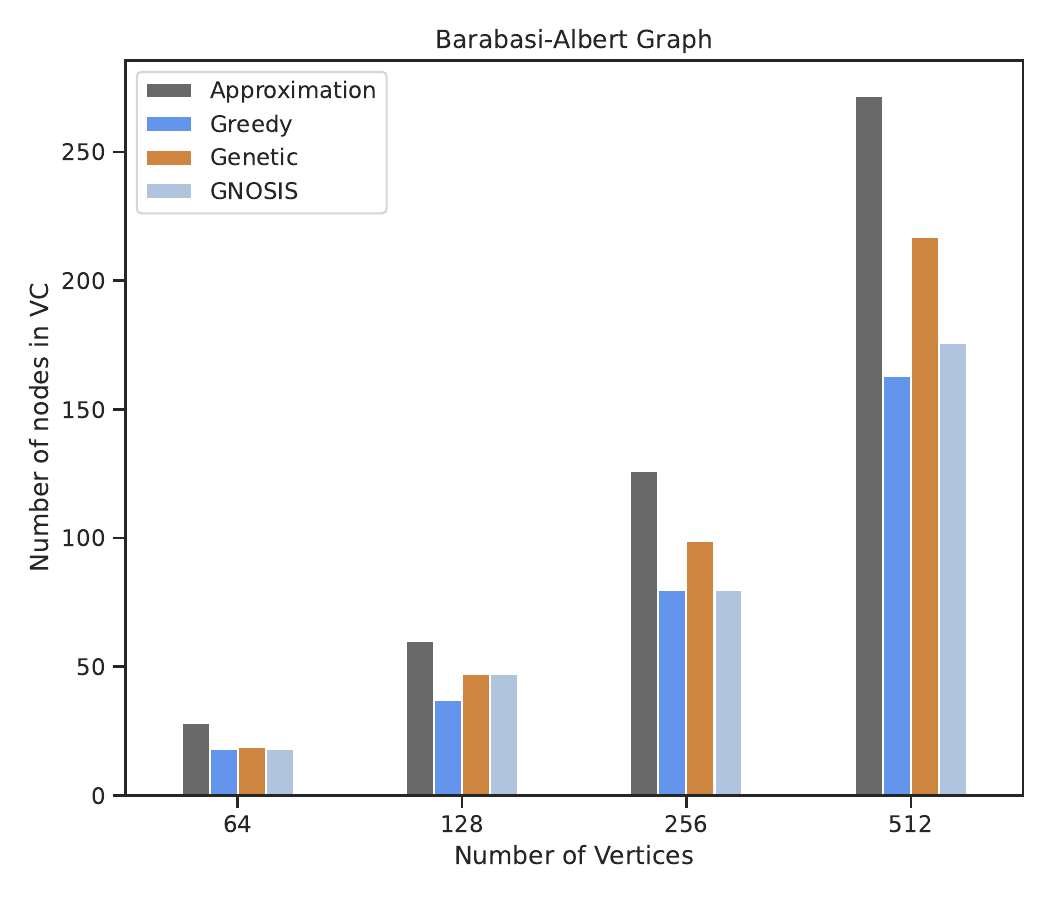}%
\label{VC_BA}}
\caption{Vertex Cover set of each algorithm for the different network topologies}
\label{vertex_cover_all}
\end{figure*}

\textit{Vertex Cover Set size analysis.}  Figure \ref{vertex_cover_all} evaluates the vertex cover set produced by each algorithm for the different network topologies. As expected, the vertex cover set's size increases in a linear fashion with an increase in the number of vertices.
As the results suggest, the Approximation algorithm generates the largest vertex cover sets, followed by the Genetic algorithm for all network topologies. 
In fact, the difference is particularly notable in the case of Watts-Strogatz and Barab{\'a}si-Albert graphs. As an example, when considering Watts-Strogatz graphs with $512$ vertices, the Approximation algorithm generates a vertex cover set size of $512$, while for Greedy, Genetic and GNOSIS, the sizes are $290$, $330$ and $282$, respectively.
In addition, for Erd\"o–R{\'e}nyi and Watts-Strogatz graphs, GNOSIS produces smaller vertex cover sets compared to the other algorithms. Only for Barab{\'a}si-Albert, GNOSIS generates a vertex cover set that is slightly larger than that produced by the Greedy algorithm.
Overall, as the results demonstrate, the Greedy algorithm competes well with the GNOSIS algorithm by producing small vertex cover sets.
On the other hand, for Barab{\'a}si-Albert graphs, GNOSIS generates the smallest vertex cover set among all the considered topologies. 
As an illustration, when considering Barab{\'a}si-Albert graphs with $512$ vertices, the GNOSIS algorithm generates a vertex cover set size of $176$ while for Erd\"o–R{\'e}nyi and Watts-Strogatz graphs, the sizes are $477$ and $288$, respectively.
Overall, for Barab{\'a}si-Albert graphs, all algorithms yield the smallest vertex cover sets compared to the other network topologies, followed by Watts-Strogatz. 

\textit{Discussion.} 
In what follows, a brief overview of the experimental simulation results and key insights is provided to summarize the performance evaluation of each algorithm.

As the results suggested, the GNOSIS algorithm exhibits higher execution times compared to the other algorithms when utilized on Erd\"o–R{\'e}nyi graphs. However, for the remaining topologies, it comes second in terms of execution time after the Genetic algorithm. Overall, the Genetic and GNOSIS algorithm manifests prolonged execution times. The reason is that GNOSIS involves training a neural network on the graph data and making predictions for each vertex in the graph. As the graph structure becomes more complex, GNOSIS captures more intricate relationships between the vertices which can lead to more accurate solutions but it also requires more computation. On the other hand, the Genetic algorithm involves the process of generating a population of potential solutions, applying genetic operators (such as crossover and mutation), and selecting the best solutions based on fitness criteria. As the size of the problem and the population increases, the time required to generate and evaluate each generation of candidate solutions also increases, resulting in longer execution time \cite{javadi2005hybrid}. In all network topologies, both Approximation and Greedy algorithms exhibit the shortest execution times, with noticeable differences when compared to other algorithms. Furthermore, as the number of vertices increases, there are no significant fluctuations between the execution time of these two algorithms.

The high execution times presented by the GNOSIS algorithm result in higher cost function values for Barab{\'a}si-Albert graphs in comparison to the other algorithms. However, GNOSIS algorithm yields lower cost function values for the remaining network topologies except for the Approximation algorithm. Moreover, the Genetic algorithm exhibits high cost function values, which are particularly noticeable in the case of Erd\"o–R{\'e}nyi graphs. On the other hand, in most cases, the Greedy algorithm yields lower cost function values compared to the Genetic algorithm, yet higher than the GNOSIS algorithm (except for Barab{\'a}si-Albert graphs). 

Although the Approximation algorithm presents the lowest cost function values and execution times in all models, it exhibits the largest vertex cover sets in all network topologies. This practically means that this algorithm is able to produce a solution faster than the other algorithms; however, this solution will contain a significantly larger amount of nodes. Hence, there exists a tradeoff between the total time taken by each algorithm to generate a solution and the number of nodes considered in the graph.
The GNOSIS algorithm produces larger vertex cover sets than the Greedy algorithm, yet smaller than the Genetic algorithm for Barab{\'a}si-Albert graphs.
However, similar to the cost function (except for the Approximation algorithm), the GNOSIS algorithm generates smaller vertex cover set sizes for the other network topologies. Genetic algorithm shows higher results compared to the Greedy algorithm for all network topologies.

To summarize, among the algorithms studied, the GNOSIS algorithm is one of the most efficient algorithms, producing the best results in terms of cost function and vertex cover set size for Erd\"o–R{\'e}nyi and Watts-Strogatz graphs. However, it is accompanied by prolonged execution times.
Additionally, the Greedy algorithm demonstrates relatively short execution times, but it ranks second after GNOSIS in terms of cost function and vertex cover set size, with the exception of Barab{\'a}si-Albert graphs.
The Genetic algorithm is characterized by high execution times across all network topologies and results in high cost function values and vertex cover set sizes, making it one of the least efficient solutions.
Finally, although the Approximation algorithm demonstrates the lowest execution times and cost function values, it is associated with the largest vertex cover set sizes, which make it unsuitable for the MVC problem.

Table \ref{tab:summary_of_all_results} provides a detailed summary of the results based on the different evaluation metrics, utilizing four tiers-Highest, Moderate, Mid, and Lowest-to illustrate the scale of values within the data.

\begin{table}[]
\centering
\footnotesize	
\caption{Summary of results}
\label{tab:summary_of_all_results}
\begin{tabular}{@{}lllll@{}}
Network Model                           & Algorithm     & ExT             & CF              & VCS             \\ \midrule
\cellcolor[HTML]{C0C0C0}Erdos-Renyi     & GNOSIS        & Highest         & Mid    & \textbf{Lowest}          \\
                                        & Approximation & \textbf{Lowest}          & \textbf{Lowest}          & Highest         \\
                                        & Greedy        & Mid& Moderate& Mid\\
\multirow{-2}{*}{}                      & Genetic        & Moderate& Highest         & Moderate\\ \cmidrule(l){2-5} 
\cellcolor[HTML]{C0C0C0}Watts-Strogatz  & GNOSIS        & Moderate& Mid& \textbf{Lowest}          \\
                                        & Approximation & \textbf{Lowest}          & \textbf{Lowest}          & Highest         \\
                                        & Greedy        & Mid& Moderate& Mid\\
\multirow{-2}{*}{}                      & Genetic        & Highest& Highest         & Moderate\\ \cmidrule(l){2-5} 
\cellcolor[HTML]{C0C0C0}Barabasi-Albert & GNOSIS        & Moderate& Highest         & Mid\\
                                        & Approximation & \textbf{Lowest}           & \textbf{Lowest}          & Highest         \\
                                        & Greedy        & Moderate        & Mid& \textbf{Lowest}          \\
\multirow{-2}{*}{}                      & Genetic        & Highest& Moderate& Moderate\\ \bottomrule
\end{tabular}
\end{table}

\begin{table*}[]
\centering
\footnotesize
\caption{Detailed results for all variations of each model}
\label{tab:detailed_resultsall}
\begin{tabular}{@{}lccccccccccccccc@{}}
\toprule
\multicolumn{4}{c}{}                                                                                                                                                                                                                                                & \multicolumn{3}{c}{\textbf{Approximation}}                                                                                                                                                                                                                                                    & \multicolumn{3}{c}{\textbf{Greedy}}                                                                                                                                                                                                                                                              & \multicolumn{3}{c}{\textbf{Genetic}}                                                                                                                                                                                                                                                                  & \multicolumn{3}{c}{\textbf{GNOSIS}}                                                                                                                                                                                                                                        \\ \midrule
\multicolumn{2}{c}{}                                                                                   & \textit{V}                           & \multicolumn{1}{c|}{\textit{E}}                                                                                     & ExT                                                                                   & CF                                                                                             & \multicolumn{1}{c|}{VCS}                                                                             & ExT                                                                                   & CF                                                                                                & \multicolumn{1}{c|}{VCS}                                                                             & ExT                                                                                     & CF                                                                                                   & \multicolumn{1}{c|}{VCS}                                                                             & ExT                                                                                     & CF                                                                                             & VCS                                                                             \\ \midrule
                     &                                                                                 & \textit{64}                          & \multicolumn{1}{c|}{\textit{\begin{tabular}[c]{@{}c@{}}63\\ 183\\ 448\end{tabular}}}                                & \begin{tabular}[c]{@{}c@{}}0.021\\ 0.020\\ 0.023\end{tabular}                         & \begin{tabular}[c]{@{}c@{}}3752.51\\ 1498.54\\ 977.61\end{tabular}                             & \multicolumn{1}{c|}{\begin{tabular}[c]{@{}c@{}}28\\ 48\\ 56\end{tabular}}                            & \begin{tabular}[c]{@{}c@{}}0.024\\ 0.026\\ 0.025\end{tabular}                         & \begin{tabular}[c]{@{}c@{}}4732.42\\ 2519.53\\ 2087.64\end{tabular}                               & \multicolumn{1}{c|}{\begin{tabular}[c]{@{}c@{}}18\\ 34\\ 45\end{tabular}}                            & \begin{tabular}[c]{@{}c@{}}1.366\\ 0.494\\ 0.423\end{tabular}                           & \begin{tabular}[c]{@{}c@{}}4734.80\\ 2702.77\\ 3829.13\end{tabular}                                  & \multicolumn{1}{c|}{\begin{tabular}[c]{@{}c@{}}19\\ 42\\ 52\end{tabular}}                            & \begin{tabular}[c]{@{}c@{}}0.24\\ 0.88\\ 1.08\end{tabular}                              & \begin{tabular}[c]{@{}c@{}}6186.54\\ 2422\\ 2365.54\end{tabular}                               & \begin{tabular}[c]{@{}c@{}}18\\ 35\\ 35\end{tabular}                            \\
                     &                                                                                 & \cellcolor[HTML]{C0C0C0}\textit{128} & \multicolumn{1}{c|}{\cellcolor[HTML]{C0C0C0}\textit{\begin{tabular}[c]{@{}c@{}}127\\ 375\\ 960\end{tabular}}}       & \cellcolor[HTML]{C0C0C0}\begin{tabular}[c]{@{}c@{}}0.041\\ 0.040\\ 0.043\end{tabular} & \cellcolor[HTML]{C0C0C0}\begin{tabular}[c]{@{}c@{}}6217.07\\ 3446.53\\ 3200.37\end{tabular}    & \multicolumn{1}{c|}{\cellcolor[HTML]{C0C0C0}\begin{tabular}[c]{@{}c@{}}60\\ 92\\ 108\end{tabular}}   & \cellcolor[HTML]{C0C0C0}\begin{tabular}[c]{@{}c@{}}0.050\\ 0.043\\ 0.063\end{tabular} & \cellcolor[HTML]{C0C0C0}\begin{tabular}[c]{@{}c@{}}8244.56\\ 5974.57\\ 6119.13\end{tabular}       & \multicolumn{1}{c|}{\cellcolor[HTML]{C0C0C0}\begin{tabular}[c]{@{}c@{}}37\\ 66\\ 89\end{tabular}}    & \cellcolor[HTML]{C0C0C0}\begin{tabular}[c]{@{}c@{}}7.533\\ 7.554\\ 1.215\end{tabular}   & \cellcolor[HTML]{C0C0C0}\begin{tabular}[c]{@{}c@{}}8393.66\\ 7554.51\\ 12317.88\end{tabular}         & \multicolumn{1}{c|}{\cellcolor[HTML]{C0C0C0}\begin{tabular}[c]{@{}c@{}}47\\ 83\\ 103\end{tabular}}   & \cellcolor[HTML]{C0C0C0}\begin{tabular}[c]{@{}c@{}}0.73\\ 4.9\\ 6.67\end{tabular}       & \cellcolor[HTML]{C0C0C0}\begin{tabular}[c]{@{}c@{}}9987.54\\ 5421.32\\ 4835.95\end{tabular}    & \cellcolor[HTML]{C0C0C0}\begin{tabular}[c]{@{}c@{}}47\\ 75\\ 76\end{tabular}    \\
                     &                                                                                 & \textit{256}                         & \multicolumn{1}{c|}{\textit{\begin{tabular}[c]{@{}c@{}}255\\ 759\\ 1984\end{tabular}}}                              & \begin{tabular}[c]{@{}c@{}}0.122\\ 0.118\\ 0.105\end{tabular}                         & \begin{tabular}[c]{@{}c@{}}12096.63\\ 9991.89\\ 9175.27\end{tabular}                           & \multicolumn{1}{c|}{\begin{tabular}[c]{@{}c@{}}127\\ 176\\ 228\end{tabular}}                         & \begin{tabular}[c]{@{}c@{}}0.142\\ 0.127\\ 0.194\end{tabular}                         & \begin{tabular}[c]{@{}c@{}}16346.34\\ 15948.10\\ 23828.37\end{tabular}                            & \multicolumn{1}{c|}{\begin{tabular}[c]{@{}c@{}}80\\ 127\\ 182\end{tabular}}                          & \begin{tabular}[c]{@{}c@{}}22.392\\ 7.621\\ 3.439\end{tabular}                          & \begin{tabular}[c]{@{}c@{}}17062.14\\ 20213.48\\ 54844.29\end{tabular}                               & \multicolumn{1}{c|}{\begin{tabular}[c]{@{}c@{}}99\\ 146\\ 202\end{tabular}}                          & \begin{tabular}[c]{@{}c@{}}1.86\\ 6.69\\ 6.5\end{tabular}                               & \begin{tabular}[c]{@{}c@{}}22775\\ 9859.43\\ 9603\end{tabular}                                 & \begin{tabular}[c]{@{}c@{}}80\\ 177\\ 180\end{tabular}                          \\
\multirow{-10}{*}{BA} & \multirow{-10}{*}{\begin{tabular}[c]{@{}c@{}}m=1\\ m=3\\ m=8\end{tabular}}       & \cellcolor[HTML]{C0C0C0}\textit{512} & \multicolumn{1}{c|}{\cellcolor[HTML]{C0C0C0}\textit{\begin{tabular}[c]{@{}c@{}}511\\ 1527\\ 4032\end{tabular}}}     & \cellcolor[HTML]{C0C0C0}\begin{tabular}[c]{@{}c@{}}0.468\\ 0.406\\ 0.478\end{tabular} & \cellcolor[HTML]{C0C0C0}\begin{tabular}[c]{@{}c@{}}25169.55\\ 33823.91\\ 47196.73\end{tabular} & \multicolumn{1}{c|}{\cellcolor[HTML]{C0C0C0}\begin{tabular}[c]{@{}c@{}}272\\ 379\\ 440\end{tabular}} & \cellcolor[HTML]{C0C0C0}\begin{tabular}[c]{@{}c@{}}0.509\\ 0.460\\ 0.819\end{tabular} & \cellcolor[HTML]{C0C0C0}\begin{tabular}[c]{@{}c@{}}36592.49\\ 61257.26\\ 100365.45\end{tabular}   & \multicolumn{1}{c|}{\cellcolor[HTML]{C0C0C0}\begin{tabular}[c]{@{}c@{}}164\\ 264\\ 358\end{tabular}} & \cellcolor[HTML]{C0C0C0}\begin{tabular}[c]{@{}c@{}}87.12\\ 28.48\\ 11.742\end{tabular}  & \cellcolor[HTML]{C0C0C0}\begin{tabular}[c]{@{}c@{}}39449.04\\ 79522.86\\ 233526.26\end{tabular}      & \multicolumn{1}{c|}{\cellcolor[HTML]{C0C0C0}\begin{tabular}[c]{@{}c@{}}217\\ 330\\ 392\end{tabular}} & \cellcolor[HTML]{C0C0C0}\begin{tabular}[c]{@{}c@{}}5.9\\ 18.43\\ 18.9\end{tabular}      & \cellcolor[HTML]{C0C0C0}\begin{tabular}[c]{@{}c@{}}51130\\ 40930.19\\ 41447.34\end{tabular}    & \cellcolor[HTML]{C0C0C0}\begin{tabular}[c]{@{}c@{}}176\\ 350\\ 345\end{tabular} \\ \midrule
                     &                                                                                 & \textit{64}                          & \multicolumn{1}{c|}{\textit{\begin{tabular}[c]{@{}c@{}}394\\ 1003\\ 1417\end{tabular}}}                             & \begin{tabular}[c]{@{}c@{}}0.021\\ 0.027\\ 0.012\end{tabular}                         & \begin{tabular}[c]{@{}c@{}}307.76\\ 385.08\\ 518.44\end{tabular}                               & \multicolumn{1}{c|}{\begin{tabular}[c]{@{}c@{}}62\\ 62\\ 62\end{tabular}}                            & \begin{tabular}[c]{@{}c@{}}0.028\\ 0.036\\ 0.034\end{tabular}                         & \begin{tabular}[c]{@{}c@{}}1708.70\\ 1027.26\\ 1200.11\end{tabular}                               & \multicolumn{1}{c|}{\begin{tabular}[c]{@{}c@{}}48\\ 58\\ 59\end{tabular}}                            & \begin{tabular}[c]{@{}c@{}}0.379\\ 0.737\\ 1.462\end{tabular}                           & \begin{tabular}[c]{@{}c@{}}3498.15\\ 7651.58\\ 10101.8\end{tabular}                                  & \multicolumn{1}{c|}{\begin{tabular}[c]{@{}c@{}}55\\ 59\\ 60\end{tabular}}                            & \begin{tabular}[c]{@{}c@{}}4.42\\ 1.77\\ 1.58\end{tabular}                              & \begin{tabular}[c]{@{}c@{}}1463.23\\ 861.89\\ 513.56\end{tabular}                              & \begin{tabular}[c]{@{}c@{}}51\\ 59\\ 62\end{tabular}                            \\
                     &                                                                                 & \cellcolor[HTML]{C0C0C0}\textit{128} & \multicolumn{1}{c|}{\cellcolor[HTML]{C0C0C0}\textit{\begin{tabular}[c]{@{}c@{}}1612\\ 4054\\ 5712\end{tabular}}}    & \cellcolor[HTML]{C0C0C0}\begin{tabular}[c]{@{}c@{}}0.037\\ 0.069\\ 0.076\end{tabular} & \cellcolor[HTML]{C0C0C0}\begin{tabular}[c]{@{}c@{}}1162.52\\ 1431.88\\ 1965.96\end{tabular}    & \multicolumn{1}{c|}{\cellcolor[HTML]{C0C0C0}\begin{tabular}[c]{@{}c@{}}124\\ 126\\ 126\end{tabular}} & \cellcolor[HTML]{C0C0C0}\begin{tabular}[c]{@{}c@{}}0.073\\ 0.152\\ 0.190\end{tabular} & \cellcolor[HTML]{C0C0C0}\begin{tabular}[c]{@{}c@{}}5041.98\\ 5343.53\\ 4722.90\end{tabular}       & \multicolumn{1}{c|}{\cellcolor[HTML]{C0C0C0}\begin{tabular}[c]{@{}c@{}}109\\ 120\\ 123\end{tabular}} & \cellcolor[HTML]{C0C0C0}\begin{tabular}[c]{@{}c@{}}1.116\\ 2.325\\ 3.512\end{tabular}   & \cellcolor[HTML]{C0C0C0}\begin{tabular}[c]{@{}c@{}}23480.76\\ 68028.0\\ 95802.12\end{tabular}        & \multicolumn{1}{c|}{\cellcolor[HTML]{C0C0C0}\begin{tabular}[c]{@{}c@{}}114\\ 122\\ 124\end{tabular}} & \cellcolor[HTML]{C0C0C0}\begin{tabular}[c]{@{}c@{}}5.4\\ 6.09\\ 6.09\end{tabular}       & \cellcolor[HTML]{C0C0C0}\begin{tabular}[c]{@{}c@{}}3038.54\\ 2084.31\\ 1979.23\end{tabular}    & \cellcolor[HTML]{C0C0C0}\begin{tabular}[c]{@{}c@{}}107\\ 117\\ 118\end{tabular} \\
                     &                                                                                 & \textit{256}                         & \multicolumn{1}{c|}{\textit{\begin{tabular}[c]{@{}c@{}}6522\\ 16331\\ 22951\end{tabular}}}                          & \begin{tabular}[c]{@{}c@{}}0.193\\ 0.377\\ 0.497\end{tabular}                         & \begin{tabular}[c]{@{}c@{}}6552.64\\ 5514.58\\ 256.0\end{tabular}                              & \multicolumn{1}{c|}{\begin{tabular}[c]{@{}c@{}}250\\ 254\\ 256\end{tabular}}                         & \begin{tabular}[c]{@{}c@{}}0.349\\ 1.018\\ 1.34\end{tabular}                          & \begin{tabular}[c]{@{}c@{}}23343.71\\ 23919.62\\ 22429.10\end{tabular}                            & \multicolumn{1}{c|}{\begin{tabular}[c]{@{}c@{}}234\\ 247\\ 250\end{tabular}}                         & \begin{tabular}[c]{@{}c@{}}3.635\\ 9.323\\ 13.38\end{tabular}                           & \begin{tabular}[c]{@{}c@{}}229235.27\\ 618366.53\\ 868933.56\end{tabular}                            & \multicolumn{1}{c|}{\begin{tabular}[c]{@{}c@{}}239\\ 248\\ 252\end{tabular}}                         & \begin{tabular}[c]{@{}c@{}}16.4\\ 27.43\\ 27.43\end{tabular}                            & \begin{tabular}[c]{@{}c@{}}7720.47\\ 17568.33\\ 12323.12\end{tabular}                          & \begin{tabular}[c]{@{}c@{}}230\\ 242\\ 242\end{tabular}                         \\
\multirow{-10}{*}{ER} & \multirow{-10}{*}{\begin{tabular}[c]{@{}c@{}}p=0.2\\ p=0.5\\ p=0.7\end{tabular}} & \cellcolor[HTML]{C0C0C0}\textit{512} & \multicolumn{1}{c|}{\cellcolor[HTML]{C0C0C0}\textit{\begin{tabular}[c]{@{}c@{}}26207\\ 65574\\ 91762\end{tabular}}} & \cellcolor[HTML]{C0C0C0}\begin{tabular}[c]{@{}c@{}}1.341\\ 2.867\\ 3.526\end{tabular} & \cellcolor[HTML]{C0C0C0}\begin{tabular}[c]{@{}c@{}}8951.86\\ 21632.86\\ 30068.60\end{tabular}  & \multicolumn{1}{c|}{\cellcolor[HTML]{C0C0C0}\begin{tabular}[c]{@{}c@{}}510\\ 510\\ 510\end{tabular}} & \cellcolor[HTML]{C0C0C0}\begin{tabular}[c]{@{}c@{}}2.479\\ 7.719\\ 14.17\end{tabular} & \cellcolor[HTML]{C0C0C0}\begin{tabular}[c]{@{}c@{}}110230.25\\ 106116.31\\ 103960.13\end{tabular} & \multicolumn{1}{c|}{\cellcolor[HTML]{C0C0C0}\begin{tabular}[c]{@{}c@{}}486\\ 502\\ 505\end{tabular}} & \cellcolor[HTML]{C0C0C0}\begin{tabular}[c]{@{}c@{}}16.01\\ 41.71\\ 56.11\end{tabular}   & \cellcolor[HTML]{C0C0C0}\begin{tabular}[c]{@{}c@{}}1942119.08\\ 5165041.22\\ 7271924.85\end{tabular} & \multicolumn{1}{c|}{\cellcolor[HTML]{C0C0C0}\begin{tabular}[c]{@{}c@{}}490\\ 501\\ 507\end{tabular}} & \cellcolor[HTML]{C0C0C0}\begin{tabular}[c]{@{}c@{}}82.43\\ 126.43\\ 123.43\end{tabular} & \cellcolor[HTML]{C0C0C0}\begin{tabular}[c]{@{}c@{}}37865.43\\ 63837.23\\ 64940.23\end{tabular} & \cellcolor[HTML]{C0C0C0}\begin{tabular}[c]{@{}c@{}}477\\ 496\\ 486\end{tabular} \\ \midrule
                     &                                                                                 & \textit{64}                          & \multicolumn{1}{c|}{\textit{\begin{tabular}[c]{@{}c@{}}99\\ 194\\ 287\end{tabular}}}                                & \begin{tabular}[c]{@{}c@{}}0.024\\ 0.024\\ 0.023\end{tabular}                         & \begin{tabular}[c]{@{}c@{}}64.0\\ 64.0\\ 64.0\end{tabular}                                     & \multicolumn{1}{c|}{\begin{tabular}[c]{@{}c@{}}64\\ 64\\ 64\end{tabular}}                            & \begin{tabular}[c]{@{}c@{}}0.028\\ 0.023\\ 0.027\end{tabular}                         & \begin{tabular}[c]{@{}c@{}}2821.53\\ 1442.30\\ 1570.44\end{tabular}                               & \multicolumn{1}{c|}{\begin{tabular}[c]{@{}c@{}}35\\ 47\\ 51\end{tabular}}                            & \begin{tabular}[c]{@{}c@{}}0.496\\ 0.311\\ 0.312\end{tabular}                           & \begin{tabular}[c]{@{}c@{}}2514.96\\ 3196.69\\ 3618.70\end{tabular}                                  & \multicolumn{1}{c|}{\begin{tabular}[c]{@{}c@{}}41\\ 48\\ 54\end{tabular}}                            & \begin{tabular}[c]{@{}c@{}}0.34\\ 0.5\\ 1.27\end{tabular}                               & \begin{tabular}[c]{@{}c@{}}2695.32\\ 1552.34\\ 869.21\end{tabular}                             & \begin{tabular}[c]{@{}c@{}}35\\ 47\\ 47\end{tabular}                            \\
                     &                                                                                 & \cellcolor[HTML]{C0C0C0}\textit{128} & \multicolumn{1}{c|}{\cellcolor[HTML]{C0C0C0}\textit{\begin{tabular}[c]{@{}c@{}}191\\ 389\\ 580\end{tabular}}}       & \cellcolor[HTML]{C0C0C0}\begin{tabular}[c]{@{}c@{}}0.056\\ 0.043\\ 0.047\end{tabular} & \cellcolor[HTML]{C0C0C0}\begin{tabular}[c]{@{}c@{}}128.0\\ 128.0\\ 128.0\end{tabular}          & \multicolumn{1}{c|}{\cellcolor[HTML]{C0C0C0}\begin{tabular}[c]{@{}c@{}}128\\ 128\\ 128\end{tabular}} & \cellcolor[HTML]{C0C0C0}\begin{tabular}[c]{@{}c@{}}0.050\\ 0.048\\ 0.052\end{tabular} & \cellcolor[HTML]{C0C0C0}\begin{tabular}[c]{@{}c@{}}5147.98\\ 3495.86\\ 3104.33\end{tabular}       & \multicolumn{1}{c|}{\cellcolor[HTML]{C0C0C0}\begin{tabular}[c]{@{}c@{}}72\\ 94\\ 102\end{tabular}}   & \cellcolor[HTML]{C0C0C0}\begin{tabular}[c]{@{}c@{}}1.399\\ 0.919\\ 0.800\end{tabular}   & \cellcolor[HTML]{C0C0C0}\begin{tabular}[c]{@{}c@{}}6212.21\\ 7416.55\\ 8444.13\end{tabular}          & \multicolumn{1}{c|}{\cellcolor[HTML]{C0C0C0}\begin{tabular}[c]{@{}c@{}}81\\ 99\\ 108\end{tabular}}   & \cellcolor[HTML]{C0C0C0}\begin{tabular}[c]{@{}c@{}}0.89\\ 1.7\\ 1.91\end{tabular}       & \cellcolor[HTML]{C0C0C0}\begin{tabular}[c]{@{}c@{}}4260.89\\ 2756.91\\ 2351.76\end{tabular}    & \cellcolor[HTML]{C0C0C0}\begin{tabular}[c]{@{}c@{}}70\\ 90\\ 93\end{tabular}    \\
                     &                                                                                 & \textit{256}                         & \multicolumn{1}{c|}{\textit{\begin{tabular}[c]{@{}c@{}}400\\ 764\\ 1144\end{tabular}}}                              & \begin{tabular}[c]{@{}c@{}}0.161\\ 0.116\\ 0.097\end{tabular}                         & \begin{tabular}[c]{@{}c@{}}256.0\\ 256.0\\ 256.0\end{tabular}                                  & \multicolumn{1}{c|}{\begin{tabular}[c]{@{}c@{}}256\\ 256\\ 256\end{tabular}}                         & \begin{tabular}[c]{@{}c@{}}0.162\\ 0.125\\ 0.126\end{tabular}                         & \begin{tabular}[c]{@{}c@{}}11187.54\\ 8555.45\\ 9418.7\end{tabular}                               & \multicolumn{1}{c|}{\begin{tabular}[c]{@{}c@{}}154\\ 188\\ 206\end{tabular}}                         & \begin{tabular}[c]{@{}c@{}}5.289\\ 3.008\\ 2.484\end{tabular}                           & \begin{tabular}[c]{@{}c@{}}14206.98\\ 19147.82\\ 31716.45\end{tabular}                               & \multicolumn{1}{c|}{\begin{tabular}[c]{@{}c@{}}165\\ 198\\ 209\end{tabular}}                         & \begin{tabular}[c]{@{}c@{}}2.72\\ 3.65\\ 5.2\end{tabular}                               & \begin{tabular}[c]{@{}c@{}}9455.43\\ 7249.61\\ 7515.97\end{tabular}                            & \begin{tabular}[c]{@{}c@{}}148\\ 181\\ 192\end{tabular}                         \\
\multirow{-10}{*}{WS} & \multirow{-10}{*}{\begin{tabular}[c]{@{}c@{}}k=2\\ k=4\\ k=7\end{tabular}}       & \cellcolor[HTML]{C0C0C0}\textit{512} & \multicolumn{1}{c|}{\cellcolor[HTML]{C0C0C0}\textit{\begin{tabular}[c]{@{}c@{}}763\\ 1529\\ 2303\end{tabular}}}     & \cellcolor[HTML]{C0C0C0}\begin{tabular}[c]{@{}c@{}}0.510\\ 0.434\\ 0.440\end{tabular} & \cellcolor[HTML]{C0C0C0}\begin{tabular}[c]{@{}c@{}}512.0\\ 512.0\\ 512.0\end{tabular}          & \multicolumn{1}{c|}{\cellcolor[HTML]{C0C0C0}\begin{tabular}[c]{@{}c@{}}512\\ 512\\ 512\end{tabular}} & \cellcolor[HTML]{C0C0C0}\begin{tabular}[c]{@{}c@{}}0.505\\ 0.489\\ 0.546\end{tabular} & \cellcolor[HTML]{C0C0C0}\begin{tabular}[c]{@{}c@{}}27571.52\\ 35829.82\\ 38984.10\end{tabular}    & \multicolumn{1}{c|}{\cellcolor[HTML]{C0C0C0}\begin{tabular}[c]{@{}c@{}}290\\ 368\\ 408\end{tabular}} & \cellcolor[HTML]{C0C0C0}\begin{tabular}[c]{@{}c@{}}25.899\\ 10.986\\ 8.841\end{tabular} & \cellcolor[HTML]{C0C0C0}\begin{tabular}[c]{@{}c@{}}35845.13\\ 79931.84\\ 124304.65\end{tabular}      & \multicolumn{1}{c|}{\cellcolor[HTML]{C0C0C0}\begin{tabular}[c]{@{}c@{}}330\\ 389\\ 416\end{tabular}} & \cellcolor[HTML]{C0C0C0}\begin{tabular}[c]{@{}c@{}}7.2\\ 11.24\\ 15.6\end{tabular}      & \cellcolor[HTML]{C0C0C0}\begin{tabular}[c]{@{}c@{}}23206.6\\ 26619.32\\ 29754.32\end{tabular}  & \cellcolor[HTML]{C0C0C0}\begin{tabular}[c]{@{}c@{}}282\\ 353\\ 388\end{tabular} \\ \bottomrule
\end{tabular}
\end{table*}

\section{Conclusion}
\label{conclusion}
The placement of container and VM images holds significant importance within the realm of Edge Computing. Given the dynamic nature of Edge applications, it is crucial to minimize cold startup times, and the swift download of application images becomes a fundamental requirement.
In this research paper, we approached the problem by modeling it as a Minimum Vertex Cover and introduced GNOSIS, a novel solution that combines actor-critic Reinforcement Learning with graph neural networks. We conducted evaluations of GNOSIS using various graph topologies and sizes, comparing its performance against various state-of-the-art approaches, namely Approximation, Greedy and Genetic.

The analysis of the results reveals that GNOSIS exhibits higher execution times but generally achieves superior MCV scores. The reason is that the utilization of Graph Neural Networks and actor-critic RL methods leverage the inherent structure of the graph, allowing for generalization to unseen graphs and the acquisition of effective strategies for making globally informed decisions. This empowers them to achieve enhanced performance and potentially discover optimal or near-optimal solutions, making them more effective in solving the minimum vertex cover problem.

While the results appear promising, our future plans involve conducting further evaluations of GNOSIS using different network types and comparing it against alternative approaches. Specifically, we aim to expand the range of considered topologies, such as balanced tree and mesh networks, and explore other methodologies for solving MVC, such as genetic meta-heuristics and Integer Linear Programming solutions.

\section*{Acknowledgments}
The research leading to these results received funding from the European Union's Horizon 2020 research and innovation programme under grant agreement No 101016509 (project CHARITY). The paper reflects only the authors' views. The Commission is not responsible for any use that may be made of the information it contains.


%





\ifCLASSOPTIONcaptionsoff
  \newpage
\fi






%

\begin{IEEEbiography}[{\includegraphics[width=1in,height=1.25in,clip,keepaspectratio]{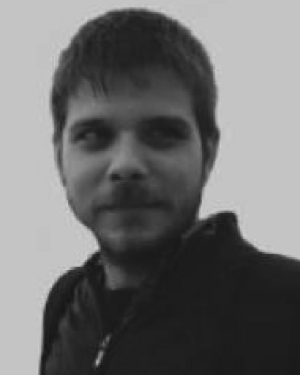}}]{Antonios Makris} is currently a Postdoctoral Researcher at the Department of Informatics and Telematics at Harokopio University of Athens and also at the School of Electrical and Computer Engineering of the National Technical University of Athens. He received his BSc Degree in Computer Science in 2013 and MSc Degree in Web Engineering in 2015, both from Harokopio University of Athens. In 2022, he received his PhD in the area of Distributed Systems from the same department. His main research interests include Distributed Computing, Edge and Cloud Computing, Big Data Management and Analysis, Machine/Deep learning, NoSQL Database Systems and Spatiotemporal and Trajectory Analysis. He has participated in numerous EU funded and national projects.
\end{IEEEbiography}
\begin{IEEEbiography}[{\includegraphics[width=1in,height=1.25in,clip,keepaspectratio]{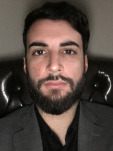}}]{Theodoros Theodoropoulos} is a Researcher at the Department of Informatics and Telematics at Harokopio University of Athens and also at the School of Electrical and Computer Engineering of the National Technical University of Athens. He is involved in several Eu-founded research projects such as Charity, Accordion and Smartship. His research interests include Deep Learning, Graph Neural Networks, Deep Reinforcement Learning, Cloud and Edge Computing.
\end{IEEEbiography}
\begin{IEEEbiography}[{\includegraphics[width=1in,height=1.25in,clip,keepaspectratio]{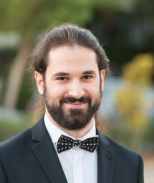}}]{Evangelos Psomakelis} is a PhD candidate at the department of Telematics and Informatics of Harokopio University of Athens and a Research Associate at the Distributed, Knowledge and Media Systems (DKMS) of the National Technical University of Athens. He is a graduate of Telematics and Informatics of the Harokopio University of Athens, holding a Master in Computer Science degree from the same university. His main research interests include machine learning, data mining and cognitive systems. He was involved in multiple European Union funded projects such as ``Consensus'', ``BASMATI'' and ``DITAS''. He is currently involved in the European funded project ``ACCORDION''.
\end{IEEEbiography}
\begin{IEEEbiography}[{\includegraphics[width=1in,height=1.25in,clip,keepaspectratio]{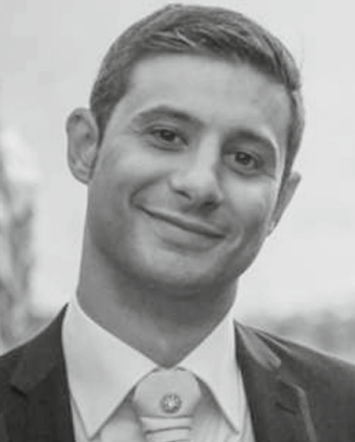}}]{Emanuele Carlini} is currently a Researcher at the Information Science and Technologies Institute (ISTI) of the National Research Council of Italy (CNR), which he joined in 2008. His main research interests revolve around many areas of distributed systems, in which he has published over 50 research papers in international and peer-reviewed journals and conferences. Currently, he is coordinating the H2020 project ACCORDION, and is active in researching distributed cloud-edge infrastructures, distributed analysis of large graphs, and AIS data analysis.
\end{IEEEbiography}
\begin{IEEEbiography}[{\includegraphics[width=1in,height=1.25in,clip,keepaspectratio]{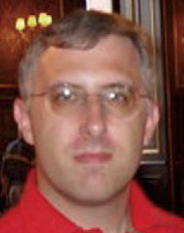}}]{Matteo Mordacchini} is currently a Researcher with the Ubiquitous Internet Lab, IIT-CNR, Italy. Currently, his main research areas include edge computing, the Internet of People paradigm, as well as adaptive, self-organizing distributed solutions. Specifically, he is investigating how models of human cognitive processes, coming from the cognitive psychology domain, can be exploited to devise autonomic and adaptive solutions for autonomous agents. His other research directions include cloud computing and opportunistic networks. He has also worked in several EU projects and has served in the TPC for many international conferences and workshops.
\end{IEEEbiography}
\begin{IEEEbiography}[{\includegraphics[width=1in,height=1.25in,clip,keepaspectratio]{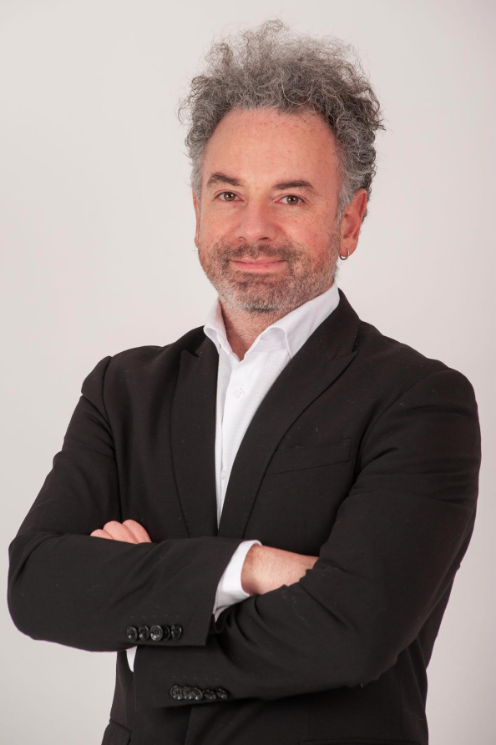}}]{Patrizio Dazzi} received his Bachelor Degree in Computer Science in 2003 and Master Degree in Computer Technologies in 2004, both from University of Pisa. In 2008 he received his Ph.D. in Computer Science and Engineering from IMT Institute for Advanced Studies Lucca, with a thesis titled ``Tools and Models for High Level Parallel and Grid Programming''. He has been a Researcher at CNR-ISTI from 2009 to 2022 and is now Senior Researcher at the Department of Computer Science of the University of Pisa. His research focuses on distributed systems, from different perspectives, ranging from high-level, programming models for adaptive applications targeting parallel and distributed environments, to cloud computing solutions and technologies with a particular interest on distributed cloud architectures and resource management on cyber-infrastructures. He is author of more than 80 research papers, the book ``Cloud Broker and Cloudlet for Workflow Scheduling'', published by Springer in the KAIST series of books. He has been organizer of workshops, conferences, journal editor and program committee member and organizer of conferences in such fields. 
He has participated to technical activities of several EC projects, including CoreGRID, GridCOMP, Contrail and NESSOS. He served as scientific coordinator of H2020 BASMATI project, as project coordinator for the European H2020 project ACCORDION and innovation coordinator of EU H2020 project TEACHING. He is a member of HiPEAC, H-Cloud and HUB4Cloud. He is co-director of the CNR-UniPi joint laboratory on Pervasive AI and former director of the ISTI-CNR node of the Italian National Laboratory on High Performance Computing. Recently, he joined the Executive Board of the IEEE Technical Committee on Cloud Computing.
\end{IEEEbiography}
\begin{IEEEbiography}[{\includegraphics[width=1in,height=1.25in,clip,keepaspectratio]{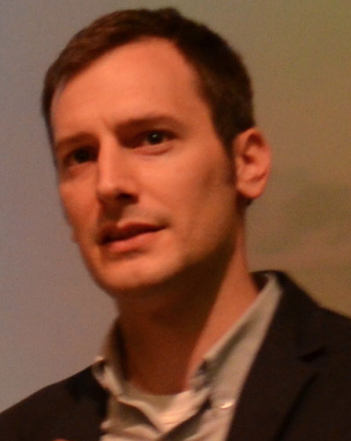}}]{Konstantinos Tserpes} is currently transitioning from a faculty position at Harokopio University of Athens to a role at the National Technical University of Athens. He holds a PhD in the area of Distributed Systems from the school of Electrical and Computer Engineering of the National Technical University of Athens (2008). His research interests revolve around efficient computing, optimizing computing systems to support novel application classes. He is working at the intersection of research areas such as cloud/edge computing, distributed systems and ML/AI. He has been involved in several EU and National funded projects leading research for solving issues related to scalability, interoperability, fault tolerance, and extensibility in application domains such as multimedia, e-governance, post-production, finance, e-health and others. He has co-authored more than 150 articles in scientific conference proceedings and journals.
\end{IEEEbiography}





\vfill


\end{document}